\documentclass[12pt]{article}             

\usepackage{ifthen}

\newboolean{ieee}

\setboolean{ieee}{false} 

\ifthenelse{\boolean{ieee}}{}{}

\ifthenelse{\boolean{ieee}}{\setlength{\textwidth}{180mm}\setlength{\textheight}{238mm}\setlength{\oddsidemargin}{-6mm}\setlength{\evensidemargin}{-6mm}
 \setlength{\headsep}{10mm}
}
{\setlength{\textwidth}{165mm}\setlength{\textheight}{229mm}\setlength{\oddsidemargin}{-1mm}
\setlength{\evensidemargin}{-4mm} \setlength{\topmargin}{-8mm}
}

\makeatletter
\renewcommand{\section}{\@startsection
{section} {1} {0mm} {-\baselineskip} {0.5\baselineskip}
{\large\bf}}
\renewcommand{\subsection}{\@startsection
{subsection} {2} {0mm} {-\baselineskip} {0.5\baselineskip}
{\normalsize\bf}} \makeatother

\usepackage{color}

\usepackage{comment} 

\excludecomment{versionC}

\usepackage{amsmath}
\usepackage{amssymb}
\usepackage{amsfonts}
\usepackage{amsthm}
\usepackage{latexsym}
\usepackage{epsfig}
\usepackage{array}
\usepackage{boxedminipage}

\newcommand{\vs}{\vspace{1mm}}
\newcommand{\vv}{\vspace{2mm}}
\newcommand{\vvv}{\vspace{3mm}}
\newcommand{\vvvv}{\vspace{4mm}}

\newcommand{\K}{\mathbb{K}}
\newcommand{\C}{\mathbb{C}}
\newcommand{\PP}{\mathbb{P}}
\newcommand{\A}{\mathbb{A}}

\newcommand{\R}{\mathbb{R}}

\newcommand{\s}{\mathbb{S}}

\newtheorem{defin}{Definition}
\newtheorem{theor}{Theorem}
\newtheorem{lema}{Lemma}
\newtheorem{propo}{Proposition}

\newtheorem{coro}{Corollary}

\newcommand{\Ma}{M[\alpha, \omega]}
\newcommand{\Mo}{M[\overline{\alpha},\overline{\omega}]}

\newcommand{\sgn}{\mathrm{sgn} \hspace{0.5mm}}
\newcommand{\rk}{\mathrm{rk} \hspace{0.5mm}}
\newcommand{\cork}{\mathrm{cork} \hspace{0.5mm}}
\newcommand{\ke}{\mathrm{ker} \hspace{0.5mm}}
\newcommand{\im}{\mathrm{im} \hspace{0.5mm}}

\newcommand{\dsp}{\displaystyle}

\newcommand{\nullity}{\mathrm{null} \hspace{0.5mm}}

\newcommand{\tra}{{\sf T}} 
\newcommand{\PQ}{\begin{pmatrix}P & Q \end{pmatrix}}

\ifthenelse{\boolean{ieee}}{\title{
Circuit Theory in Projective Space and \\
Homogeneous Circuit Models}
\author{Ricardo Riaza\thanks{
\hspace{-4mm} 
Dept.\  Matem\'{a}tica Aplicada a las TIC
\& Information Processing and Telecommunications Center, 
ETSI 
Telecomunicaci\'{o}n, Universidad Polit\'{e}cnica de Madrid, Spain. {\sl ricardo.riaza@upm.es}. Research supported by Project 
MTM2015-67396-P (MINECO/FEDER). The author acknowledges several stimulating discussions  on this topic
with
P. Ituero, M. Lambea and A. Sanz.}
}
}{\title{
Circuit Theory in Projective Space and \\
Homogeneous Circuit Models}
\author{Ricardo Riaza\thanks{
Dept.\  Matem\'{a}tica Aplicada a las TIC
\& Information Processing and Telecommunications Center, 
ETSI 
Telecomunicaci\'{o}n, Universidad Polit\'{e}cnica de Madrid, Spain. {\sl ricardo.riaza@upm.es}. Research supported by Project 
MTM2015-67396-P (MINECO/FEDER). The author acknowledges several stimulating discussions  on this topic
with
P. Ituero, M. Lambea and A. Sanz.}
}
}

\date{} 

\begin{document}

\markboth{IEEE Transactions on Circuits and Systems -- I REGULAR PAPERS
(SUBMITTED)}
{Riaza: Circuit Theory in Projective Space and
Homogeneous Circuit Models (SUBMITTED)}

\maketitle


\ifthenelse{\boolean{ieee}}{}{\mbox{}\vspace{-13mm}} 

\begin{abstract}
This paper presents a 
general framework for 
linear circuit analysis
based on elementary aspects of projective geometry.
We use a flexible approach in which no a priori assignment of
an electrical nature to the circuit branches is necessary.
Such an assignment is eventually done just by setting certain 
model parameters, in a way which avoids 
the need for a distinction between voltage and current sources
and, additionally, 
makes it 
possible to get rid of voltage- or current-control
assumptions 
on the impedances. This 
paves the way for
a completely general $m$-dimensional
reduction of any 
circuit defined by $m$ two-terminal, uncoupled linear elements,
contrary to most 
classical methods which at one
step or another impose certain restrictions on the allowed
devices.
The reduction has 
the form
$$\begin{pmatrix} AP \\ BQ \end{pmatrix} u =
\begin{pmatrix} AQ \\ -BP \end{pmatrix} \bar{s}.$$
Here, $A$ and $B$ 
capture the graph topology, 
whereas $P$, $Q$, $\bar{s}$ comprise 
homogeneous descriptions of all the circuit elements;
the unknown
$u$ is an $m$-dimensional vector of (say) ``seed'' variables from which
currents and voltages are obtained 
as
$i=Pu -Q
\bar{s}$, \hspace{0.5mm}$v=Qu + P
\bar{s}$. 
Computational implementations
are straightforward. 
These models
allow for a general characterization of non-degenerate
configurations in terms of the multihomogeneous
Kirchhoff polynomial, and in this direction we present some
results of independent interest involving the matrix-tree theorem.
Our approach can be 
easily
combined with 
classical 
methods
by using homogeneous descriptions
only for certain branches, yielding partially homogeneous models. 
We also indicate 
how to accommodate controlled sources and coupled devices 
in the homogeneous framework.
Several examples illustrate the results.
\vspace{-3mm}
\end{abstract}

\ifthenelse{\boolean{ieee}}{}{\newpage} 

\begin{versionC}
\color{red}

\begin{center}

{\Large\bf OLD: Symmetric reductions and universal objects \vspace{2mm}\\
in electrical circuit theory\footnote{Supported by Research Project 
MTM2015-67396-P
of Ministerio de Econom\'{\i}a y Competitividad, Spain/FEDER.}}

\vspace{6mm}

{\sc 
Ricardo Riaza}\\ 
\ \vspace{-3mm} \\
Depto.\ de 
Matem\'{a}tica Aplicada 
a las TIC, 
ETSI 
Telecomunicaci\'{o}n  \\ 
Universidad Polit\'{e}cnica de Madrid -  
28040 Madrid, Spain \\
{\sl ricardo.riaza@upm.es} \\



\end{center}


\begin{abstract}

Self-contained (projective stuff)

\

(ABSTRACT OLD)

Non-passive

Linear or not

Uncoupled

\

Emphasize global

Fully nonlinear

Core variables: in case there exists a global description of a given characteristic in terms
of a physical variable (current, voltage, charge, flux), then the corresponding core variable
can be identified with this physical one at no cost (e.g.\ for globally current-controlled resistors
their current qualifies as a core variable). The formalism accounts automatically for that ($\psi=id$ etc)
and only the physical meaning of the variable is lost (at first sight).
But e.g. a hysteresis loop precludes a global description in terms of a physical controlling variable,
contrary to the core variable which accounts also for that.

Even in the presence of global physical explicit descriptions [abreviar esta expresi\'on] our
formalism is proved useful. E.g. in linear cases el tema v=Ri excluye G=0 in the description of 
degenerate configurations. This idea is deeper than it seems; actually a universal polynomial
from which all (say, affine) tree polynomials (Kirchhoff, Maxwell, hybrid) can be recovered.

\end{abstract}



\vspace{2mm}




\newpage


\begin{center}
{\large\bf Circuit theory in projective space}
{\large\bf and}
{\large\bf homogeneous circuit models}
\end{center}


\

\vv

\noindent \begin{boxedminipage}{17cm}

\vv

2b. Sources. Graph + configurations. LTI circuit: graph
with an assignment of $\PP \times {\cal F}$ weight
Think equivalence relation
$(V \times {\cal F})/\sim$ con $V=\R^2-\{0\}$). Particular DC circuits,
AC circuits.

\vv

\vv

Future: companion paper nonlinear version. Later on: coupled problems,
controlled sources, multiterminal devices. Non-smooth problems. Distributed systems. Etc.

\end{boxedminipage}

\

\newpage

``Big goals'' or advantages of the projective approach:

\

1. Universal polynomial characterizing all regular configurations; whole set of parameter values characterizing
existence and uniqueness of solutions without controlling assumptions.

\vv

1.1 Of independent interest: projective version of the matrix-tree theorem; maybe known (or can be derived
from known from known results) in the context of
matroid theory but at least not so in circuit theory (Chen's book).

\

2. Framework to accommodate circuit modelling techniques in a comprehensive manner:

\vv

2.1 Spaces involved, reduced circuit descriptions: where they are well-defined, why... 
E.g. tree-based reductions.

\vv

2.2 Characteristic reduction (universal); nonlinear
version, get rid of voltage- or current-controlled assumptions. 

\vv

2.3 Generalized nodal reduction.

\newpage
\color{black}
\end{versionC}

\section{Introduction}
\label{sec-intro}

In order to motivate the ideas discussed in this paper, 
let us consider a seemingly elementary analytical problem. Assume 
we are given a loop composed of three impedances (see Fig.\ 2(a); 
we will return to this example in Section \ref{sec-homog}) 
and that we are interested
in characterizing the parameter (impedance) values 
for which the circuit is non-degenerate, i.e.\ has a unique solution.
We allow 
the real part of the impedances to become zero or
negative since otherwise the problem is trivial.  

A tentative answer is that the non-degeneracies are defined by the condition
$z_1 + z_2 + z_3 \neq 0$, because otherwise any
non-trivial loop current yields a solution. But it may come as a little 
surprise that the vanishing of $z_1 + z_2 + z_3$ does not describe
{\em all} possible degeneracies. Indeed, we are presuming here
that all three devices admit a description in terms of 
the impedance parameters $z_k$ ($k=1,$ $2,$ $3$), ruling out open-circuits. If we 
use instead admittance descriptions, via e.g.\ nodal analysis one can check
that the degeneracy condition is defined by the vanishing of the
polynomial $y_1y_2 + y_1 y_3 + y_2y_3$; but short-circuits are now excluded. 
Combining admittance and impedance
descriptions for different devices 
we would get hybrid
polynomials characterizing degenerate 
configurations in other contexts, but never including all
possible scenarios.

One of our goals is to have a single (and simple) description of all
possible degenerate parameter sets. This can be achieved by 
resorting to a {\em homogeneous}
description of the devices: instead of using
the impedance  $z_k$ or the
admittance $y_k$, we describe each device in terms of a pair
of parameters
$(p_k:q_k)$ not vanishing simultaneously. If $q_k \neq 0$
then the admittance $y_k$ is well-defined as the quotient $p_k/q_k$;
conversely, when $p_k \neq 0$  the impedance $z_k$ is $q_k/p_k$. 
The pair $(p_k:q_k)$ now captures all possible cases; this
is of interest e.g.\ when one wishes to include both short- and 
open-circuits in the same model, for instance in the computation
of Th\'evenin equivalents or in the presence of ideal switches. In
the homogeneous setting both $z_1 + z_2 + z_3$ and
 $y_1y_2 + y_1 y_3 + y_2y_3$ (among other polynomials) will arise as 
so-called dehomogenizations
of a single universal polynomial (cf.\ (\ref{kir-hom})), namely
 $p_1p_2q_3 + p_1 q_2 p_3 + q_1 p_2 p_3$ in our present case, whose zeros
describe {\em all} degeneracies.

The homogeneous approach provides a key advantage in circuit analysis
when we drive the idea further. Specifically, we will
describe the characteristic of each device 
in terms of 
just one variable $u_k$ (to be called a
{\em homogeneous} variable)
by means of the parametric form of Ohm's law
defined by the relations 
$i_k=p_k u_k$, $v_k = q_k u_k$.
Combining this 
with
Kirchhoff laws we avoid 
the need
to assume specific impedance- or admittance-descriptions for the devices,
and arrive at a general family of reduced circuit models
comprising all possible parameter settings. Sources are easily accommodated,
as shown later.
The approach can be also naturally combined
with classical analysis methods by using homogeneous descriptions
only for certain branches.

\begin{versionC}
\color{red}
In order to motivate the ideas to be discussed in this paper, 
let us consider two (seemingly elementary) analytical problems.
Assume we are interested in measuring, or simulating in a computer, the open-circuit voltage
and the short-circuit current across two terminals of a given circuit,
typically to compute its Th\'evenin equivalent.
To simplify 
matters as much as possible, let us 
consider the (possibly) simplest setting for this, defined by the circuit in Fig.\ 1(a). 


\begin{figure}[ht] \label{fig-thevenin}
\vspace{2mm}
\ifthenelse{\boolean{ieee}}{\hspace{-2mm}}{\hspace{15mm}} 
\parbox{0.5in}{
\ifthenelse{\boolean{ieee}}{\epsfig{figure=thevenin.eps, width=0.25\textwidth}}
{\epsfig{figure=thevenin.eps, width=0.3\textwidth}}
\ifthenelse{\boolean{ieee}}{
\put(-95,57){\small $v_0$}
\put(-104,70){\footnotesize $+$}
\put(-72.5,96){\small $z_1$}
\put(2,70){\footnotesize $+$}
\put(3,56){\small $v$}
\put(2,41){\footnotesize $-$}
\put(-22,106){\small $i$}}
{\put(-105,62){\small $v_0$}
\put(-111,75){\footnotesize $+$}
\put(-79.8,105){\small $z_1$}
\put(3,77){\footnotesize $+$}
\put(4,61){\small $v$}
\put(3,44.5){\footnotesize $-$}
\put(-24.5,117){\small $i$}
}}
\ifthenelse{\boolean{ieee}}{\hspace{39.5mm}}{\hspace{80mm}}
\parbox{0.5in}{\vspace{0mm}
\ifthenelse{\boolean{ieee}}{\epsfig{figure=tank.eps, width=0.1635\textwidth}}{\epsfig{figure=tank.eps, width=0.19\textwidth}}
\ifthenelse{\boolean{ieee}}{
\put(-79,39){\small $z_1$}
\put(-13,39){\small $z_2$}
}
{
\put(-84,41){\small $z_1$}
\put(-14.3,41){\small $z_2$}
}}
\ifthenelse{\boolean{ieee}}{\caption{(a) A voltage divider. \hspace{13mm} (b) Short-circuiting the source.}}{\caption{(a) A voltage divider. \hspace{32mm} (b) Short-circuiting the source.\hspace{3mm}}}
\end{figure}

The dotted box is just aimed to indicate
that, in essence, the same remarks
would apply for whatever circuit there might be inside it.
 Elementary circuit theory yields for the voltage divider the relation 
\begin{equation}\label{thevenin1}
z_1i + v=v_0.  
\end{equation} 
As a circuit model,
(\ref{thevenin1}) is however underdetermined since no relation between
the output variables $i$ and $v$ is yet specified. 
If we assume a voltage-controlled form for the load 
and write 
accordingly such relation as $i = y_2v$, then (\ref{thevenin1})
reads as
\begin{equation}\label{thevenin2}
(z_1y_2 + 1)v=v_0.
\end{equation} 
In particular, we model an open-circuit here by choosing $y_2=0$, yielding the open-circuit
(Th\'evenin) voltage,
which of course equals $v_0$ in our simplified context;
by contrast, the dual 
(short-circuit) case is not smoothly accommodated in (\ref{thevenin2}). 
Conversely, if we write $v = z_2 i$ then (\ref{thevenin1}) amounts
to 
\begin{equation}\label{thevenin3}
(z_1 + z_2)i = v_0.
\end{equation} 
Now  a short-circuit is captured by setting $z_2=0$ and this allows
one to derive the short-circuit (Norton) current; 
but there is no chance to
describe here the 
open-circuit 
case.

With \eqref{thevenin2} and \eqref{thevenin3}  we reflect 
the fact that, 
in practice,
two different models are usually set up
to compute the Th\'evenin voltage
and the Norton current (and subsequently the Th\'evenin impedance)
of 
any given circuit at a 
pair of terminals. Our first problem
is to formulate a one-equation model on a single 
indeterminate (such as (\ref{thevenin2}) or (\ref{thevenin3})),
but capturing all possible cases for the output;
in particular our model should accommodate simultaneously the open- and
short-circuit cases by adjusting certain circuit parameters, 
something that as indicated above 
neither (\ref{thevenin2}) nor (\ref{thevenin3}) do. 

The second problem, closely related to the one above, 
seeks to characterize the whole
set of non-degenerate pairs of values for both impedances, 
that is, the set of parameter values defining circuits with 
a unique solution.
It is easy to realize that this does not depend
on the actual value of $v_0$: therefore, we are allowed to fix $v_0=0$
and analyze
the  circuit depicted in 
Fig.\ 1(b). 
A tentative answer is
that the non-degeneracy condition is $z_1 + z_2 \neq 0$, since
otherwise any non-trivial loop current solves the circuit 
of Fig.\ 1(b)
(this
can be also derived by looking at the coefficient of (\ref{thevenin3})).
But this is only partially true; indeed, 
if we assume that the load has a
voltage-controlled description in terms of an admittance $y_2$ 
then the condition can be checked to be $z_1y_2+1 \neq 0$
(cf.\ the coefficient of (\ref{thevenin2})).
When admittance descriptions are given for both $y_1$ and $y_2$, then 
we get
$y_1 + y_2 \neq 0$; there is also a fourth setting in which
the non-degeneracy condition is $y_1z_2+1 \neq 0$. Our second problem is
to capture all non-degenerate sets of parameter values in a single 
(and simple) equation.

A solution 
to both problems comes from resorting to a {\em homogeneous}
description of the impedances. Indeed, assume that the load 
is not described by either an  admittance parameter 
$y_2$ or an 
impedance parameter 
$z_2$, but by a pair of parameters
$(p_2:q_2)$ not vanishing simultaneously. If $q_2$ does not vanish
then the admittance $y_2$ is well-defined as the quotient $p_2/q_2$;
conversely, when $p_2 \neq 0$  the impedance $z_2$ is $q_2/p_2$. Note that modelling
the load impedance by the pair $(p_2:q_2)$ captures all cases.

As detailed later, a completely general model
(not only for this one but for arbitrary circuits) can be
formulated in this framework by describing both the current $i$ and the
voltage $v$ in terms of a single abstract variable $u$ (to be called,
with a certain terminological abuse, a 
{\em homogeneous variable}) by means of the
relations  $i=p_2u$, $v = q_2u$. From (\ref{thevenin1})
we then derive the desired model
\begin{equation}\label{thevenin4}
(z_1p_2 + q_2)u=v_0.
\end{equation} 
In particular, with $p_2 =0$ we describe an open circuit at the output;
solving for $u$ in this case readily yields the Th\'evenin voltage 
in the
form $v=q_2u$, whatever $q_2\neq 0$ is: we can fix $q_2=1$ for simplicity. 
In the same model we accommodate the short-circuit case simply by
setting $q_2=0$; the corresponding solution of (\ref{thevenin4}) then
provides the Norton current as $i = p_2u$ for any non-vanishing choice of $p_2$; again $p_2=1$ is possible. The reader should not get confused by the
simplicity of the example: we emphasize that the same reasoning 
essentially holds
when the output is defined by any pair of terminals in any linear circuit.
We note in passing that there are other (closely related) models which also solve the problem.

The solution to the second problem is slightly more involved. A comprehensive 
discussion will be given later, but for the moment let
us say that using a homogeneous description also for the
first impedance characterizes the whole set of degenerate configurations
as the zero set of a (so-called {\em multihomogeneous}) polynomial
\begin{equation}\label{polyn1}
q_1p_2 + p_1 q_2.
\end{equation}
Indeed, the non-degeneracy condition $z_1 + z_2 \neq 0$ is obtained here from recasting
as $p_1 \neq 0$, $p_2 \neq 0$ the assumption that impedance or current-controlled
descriptions exist;
allowed by this, divide the polynomial (\ref{polyn1}) by $p_1p_2$ to get
$q_1/p_1 + q_2/p_2$, that is, $z_1 + z_2$. Readers familiar
with the matrix-tree theorem, a topic that will be extensively
discussed later, will realize that this is actually
the (co-tree) classical form of the Kirchhoff or tree-enumerator polynomial.
The other polynomials showing up
above, namely $y_1 + y_2$, $z_1y_2+1$ and $y_1z_2+1$, arise in a similar
way
under other assumptions on the existence of impedance/admittance descriptions. 
The former is Maxwell's (tree-based) form
of the Kirchhoff polynomial, whereas the last two arise in hybrid formulations. All of them
are dehomogenizations of the universal polynomial (\ref{polyn1}). We may even dehomogenize 
only some of the variables: for instance, divide (\ref{polyn1}) by $p_1$ and write 
$q_1/p_1=z_1$ to obtain the coefficient in the left-hand side of (\ref{thevenin4}). 


\color{black}
\end{versionC}

Using homogeneous descriptions is not new in circuit theory;
we can cite, at least, the works \cite{OtroBryantProjective,
chaiken, chenFilters, penin, smith72}. 
However, from the point
of view of the author this formalism has not been 
systematically exploited or even fully developed. 
In this direction, the main goal of this paper is to
present a comprehensive framework (based on elementary aspects of
projective geometry)
accommodating homogeneous descriptions of linear circuits
and then, allowed by the broad generality of this approach,
to formulate new analysis methods 
and address certain analytical aspects of circuit theory.

Specifically, we
will perform a construction in which any
linear 
circuit, typically in a resistive
DC context or in AC sinusoidal steady state (but also in the Laplace
domain), can be modelled
as a directed graph whose branches 
are endowed with weights
from a certain subset of a 
projective plane, in a way such
that branches need not be classified {\em a priori} as 
sources or impedances/resistors. 
This will make
it possible to get rid of voltage/current control assumptions
in the formulation of reduced 
models. 
This approach
also accounts for
the Th\'evenin/Norton duality when modelling non-ideal 
sources and, at a higher
abstraction level, in the formulation of equivalent circuits.
In this setting, 
from any given digraph a particular linear circuit is defined when
choosing a specific {\em configuration,} that is, a set
of homogeneous parameters for all branches.
This framework is presented in Section \ref{sec-proj}. 

The homogeneous formalism allows one to provide a precise
characterization of the conditions under which certain families of reduced
circuit models are well-defined, in terms of 
configurations. Actually, we can frame most 
analysis methods 
as families of reductions of the circuit equations to certain
subspaces of the so-called Kirchhoff and characteristic spaces. 
By using a suitable parametrization of the characteristic
space 
we introduce 
a family of homogeneous branch models
which (contrary to 
branch-voltage, branch-current and hybrid methods, but also to nodal
and loop analysis models, which at one step or another require
certain voltage/current control assumptions) 
are completely general in the sense that
they are well-defined for all possible
configurations. 
This material can be found in Section  \ref{sec-homog},
which is closed with some examples. 

In Section \ref{sec-nondeg}
we benefit from the homogeneous formalism in the
characterization, for any given digraph, 
of the set of non-degenerate configurations,
which are obtained as the complement of the zero set of a 
universal form of the Kirchhoff or tree-enumerator polynomial. 
Different dehomogenizations of this universal 
polynomial 
yield specific polynomials arising in different analysis methods.
Closely related to this are several results  of independent interest
involving the well-known
matrix-tree theorem which are discussed in that section.


We sketch the way in which this approach can be extended to
include controlled sources and coupled problems in 
Section \ref{sec-controlled}.
Some potential applications of this formalism
in fault isolation problems are discussed
in Section \ref{sec-faults}.
Finally, concluding remarks can be found in Section \ref{sec-con}. 




\section{Linear circuits as projectively weighted digraphs}
\label{sec-proj}

\subsection{Background: projective lines and planes}

There are many excellent books on projective geometry 
which may provide a reader with an introduction to this topic,
if necessary. 
We refer him/her in particular to \cite{jeangallier, semple, 
karensmith}.
Note that we will only use elementary aspects 
of this discipline and all the necessary background is compiled here.

In the sequel we assume that $\mathbb{K}$ is either
$\R$ or $\C$: the real case will be used to model resistive circuits with
DC sources, whereas 
the complex case accommodates, via phasors, circuits with AC sources in 
sinusoidal steady state and, in greater generality, can be used 
to analyze circuits in the Laplace domain.
Let $V$ be a finite-dimensional vector space over $\K$: we will
be interested in two- and three-dimensional cases. 
\begin{versionC}
\color{red} 
(this way we may accommodate
the space $E-\{0\}$ of non-trivial linear linear functions from $\K^2$ to $\K$; what matters
in Ohm's law is the kernel (characteristic) and therefore the linear form is defined
up to a constant, hence the projective formalism)
\color{black}
\end{versionC}
We define an equivalence relation in the set of non-null vectors
$V-\{0\}$ by letting
\begin{equation}\label{equiv}
u \sim v \text{ if } u = \mu v \text{ for some } \mu \in \K-\{0\}.
\end{equation}
The set of equivalence classes (to be denoted 
as $\PP(V)$) is a {\em projective space} and
each class is a projective point. When $V$ has dimension
two or three $\PP(V)$ is a {\em projective line} or a 
{\em projective plane},
respectively. Note that the dimension of $V$ exceeds by 
one the dimension of the projective space $\PP(V)$. 
In particular we denote $\PP(\K^2)$ by
$\K\PP$ and $\PP(\K^3)$ by $\K\PP^2$.




Let us fix for simplicity $V=\K^2$ in order to focus on 
the projective line, either with
$\K=\R$ or $\K=\C$ (be aware of the fact that a complex line has
complex dimension one). A point in
$\K\PP$ is an equivalence class defined by the set of non-null vectors
in $\K^2$ within any straight line through the origin. If we
take any representative $(p,q)$  of this class 
(with either $p$ or $q$, or both, 
non-null), we may describe the whole equivalence class,
that is, the corresponding point of $\K\PP$, by writing 
$(p:q)$, where
the notation is aimed to distinguish the representative
(namely, $(p,q)  \in \K^2-\{0\}$) from the whole class.
In other words, $(p_1:q_1)$ and
$(p_2:q_2)$ denote the same projective point if (and only
if) there is a non-zero constant $\mu$ such that $p_1 = \mu p_2$ and
$q_1 = \mu q_2$. Making implicit use of the canonical
basis of $\K^2$, we say that $(p:q)$ are {\em homogeneous coordinates}
for the projective point 
(we use the same notation $(p:q)$ for homogeneous coordinates
arising from the canonical basis
and for the projective point as an 
equivalence class, to avoid the cumbersome $[(p,q)]$ for the 
latter). 
Note that 
homogeneous
coordinates are defined only up to a non-vanishing constant; that is,
a point in the projective
line is not described by a unique coordinate pair but
by a whole set of such pairs.
As detailed later, homogeneous coordinates in the projective plane
$\K \PP^2$ are defined
in an entirely analogous manner; here we choose triads $(p:q:s)$ with
at least one non-vanishing entry.

Either in $\K\PP$ or in $\K\PP^2$ (in any projective space, actually), 
the non-vanishing of 
a homogeneous coordinate 
will define an 
{\em affine patch}, namely, a space which can be identified 
either with $\K$ or with $\K^2$, and which only misses from 
$\K\PP$ or from $\K\PP^2$ the point or the line 
(respectively) at infinity.
We will use this in the circuit
context to define the current-controlled (impedance) 
and the voltage-controlled (admittance)
patches, characterized by the conditions $p \neq 0$ and $q \neq 0$.

Later on we will use the above construction with 
other  vector spaces $V$, specifically
the 
space $L(\K^2, \K)$ of linear forms from $\K^2$ to $\K$ 
and the space of polynomials of degree not greater than 
one in two indeterminates 
and with coefficients in $\K$. 
Note that 
when a projective space is constructed from an arbitrary 
vector space $V$, 
homogeneous coordinates are defined
after fixing a basis in $V$ (or, more generally,
a projective frame in the projective space $\PP(V)$, 
see again \cite{jeangallier}).


\subsection{Impedance 
in the projective line}
\label{subsec-impedances}



To get a lighter exposition, from now on our terminology
focuses on the complex case; therefore, in most cases we only 
speak of impedance,
admittance, etc. and not of resistance, conductance and so on.
In any case, the notation $\K$ recalls that all results apply in
both 
the real and the complex setting.

Elementary circuit theory says 
that the characteristic of a linear circuit element
is governed by Ohm's law,
\begin{equation} \label{ohm}
v=zi,
\end{equation}
where $z$ represents 
impedance. 
This current-controlled description
formally excludes an open-circuit.
By contrast, the latter (but not a short-circuit)
is accommodated in the 
voltage-controlled or admittance description 
\begin{equation} \label{ohm2}
i=yv.
\end{equation}

We may easily accommodate all cases above 
by means of
an {\em homogenization} of either (\ref{ohm}) or (\ref{ohm2}):
namely,
setting $z=q/p$ in (\ref{ohm}) and multiplying by $p$, 
or recasting $y=p/q$ in (\ref{ohm2}) and then multiplying by $q$, we
arrive at the so-called homogeneous form of Ohm's law,
\begin{equation}\label{ohmh}
pv-qi=0,
\end{equation}
where the parameters $p$ and $q$ cannot vanish simultaneously.
This approach essentially makes a {\em projective completion}  
\cite{jeangallier} of the 
complex line where either the impedance or the admittance lies; 
focusing e.g.\ on 
(\ref{ohm}), we may rephrase the above approach by recasting
the impedance $z$ in homogeneous coordinates as $(1:z)$ 
and then allowing the first coordinate to vanish in order to accommodate
the infinite impedance case.

But we may do the same from scratch in a way which will admit
a natural extension to problems with sources and also to
the nonlinear context. Note that what
really matters in any form of Ohm's law is the set of values
of $i$ and $v$ which satisfy the device characteristic. All 
(\ref{ohm}), (\ref{ohm2}) and (\ref{ohmh}) can be understood
to define the kernel of a certain 
{\em regular} (i.e.\ not identically zero)
linear form $f:\K^2 \to \K$ capturing
this characteristic, where the domain
$\K^2$ describes the $(i,v)$-space. 
One may choose many
different ways to write this linear form, 
(\ref{ohm}) and (\ref{ohm2}) being particular instances which
are valid under certain (broad but not fully general) 
hypotheses. By contrast, 
(\ref{ohmh}) includes 
all cases. Moreover, it is obvious that 
multiplying (\ref{ohmh}) by any non-null constant $\mu$ we get
the same zero set and hence another admissible description 
of the same relation; equivalently,
the kernel of $f$ and of $\mu f$ are the same. This way the projective
formalism arises naturally.

Let then $V$ stand for the vector space of linear forms
$L(\K^2, \K)$, restrict the attention to regular forms ($f \not\equiv 0$)
 and consider the projective line $\PP(V)$
defined by the equivalence classes $[f]$ under the relation (\ref{equiv}).
This means that, 
mathematically, we look at an impedance (or at a linear resistor)
as a point in this complex (resp.\ real) projective line, that is,
as {\em an equivalence class of regular linear forms}. 
If we let $f$ be a representative in $V$
of the equivalence class of a given impedance,
and by defining the basis $f_1(i,v)=v$ and $f_2(i,v)=-i$
of $V$, the impedance is defined by the homogeneous coordinates $(p:q)$
which make $f(i,v)=pf_1(i,v)+qf_2(i,v)=pv-qi$, as in (\ref{ohmh}).
We will say that  $(p:q)$ is a {\em homogeneous description}
of the impedance (or the resistance in the real case).

Equivalently, you can identify the vector 
spaces $L(\K^2, \K)$ and $\K^2$ 
via the isomorphism 
$\alpha v + \beta i \to (\alpha, -\beta)$,
and then look at $(p:q)$ as the homogeneous 
coordinates defined by the canonical basis in $\K^2$.
From now on we assume implicitly the use of this isomorphism to denote 
by $\K\PP$ the projective line where (homogeneous) impedances lie.

In this framework,
the set of projective points defined by homogeneous coordinates
$(p:q)$ with $p \neq 0$ will be called
the  {\em current-controlled} or {\em impedance}
patch and denoted by $\mathbb{A}_{\check{z}}$.
Given any 
point
in this patch, the impedance parameter 
(or the resistance, in the real setting)
is uniquely defined from any pair of homogeneous coordinates $(p:q)$
as $z=q/p$.
The remaining point in $\K\PP-\A_{\check{z}}$, with homogeneous
coordinates $(0:q)$, can be seen as the point
of infinite impedance.
Analogously, 
the {\em voltage-controlled} or {\em admittance} 
patch $\A_{\check{y}}$ is defined by the condition $q \neq 0$;
in this case the admittance (or the conductance in the real context)
is again well defined as $y=p/q$.
Note finally that the point of infinite impedance corresponds
to the one of zero admittance, and vice-versa.


\subsection{Sources and the projective plane}
\label{subsec-sources}


The ideas in the previous subsection can be extended to accommodate
also active elements (sources) in a natural way. 
To this end, consider 
a voltage source $v_s$ in series with 
an impedance 
$z$, as in Fig.\ 1(a).
\begin{figure}[ht] \label{fig-sources}
\vspace{2mm}
\ifthenelse{\boolean{ieee}}{\hspace{1mm}}{\hspace{26mm}}
\parbox{0.5in}{
\ifthenelse{\boolean{ieee}}{\epsfig{figure=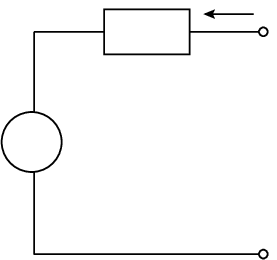, width=0.18\textwidth}}{\epsfig{figure=voltsource.eps, width=0.2\textwidth}}
\put(-68,38){\small $v_s$}
\put(-76,54){\footnotesize $+$}
\put(-45,76.5){\small $z$}
\put(5,76.4){\footnotesize $+$}
\put(6,38){\small $v$}
\put(5,0){\footnotesize $-$}
\put(-15,90){\small $i$}
}
\ifthenelse{\boolean{ieee}}{\hspace{33mm}}{\hspace{60mm}}
\parbox{0.5in}{
\ifthenelse{\boolean{ieee}}{\epsfig{figure=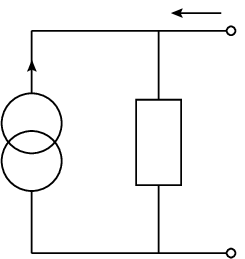, width=0.16\textwidth}}{
\epsfig{figure=currsource.eps, width=0.177\textwidth}}
\put(-66,62){\small $i_s$}
\put(-30.6,39){\small $y$}
\put(5,76.4){\footnotesize $+$}
\put(6,38){\small $v$}
\put(5,0){\footnotesize $-$}
\put(-15,90){\small $i$}
}
\vspace{2mm}
\caption{Non-ideal voltage and current sources.}
\end{figure}

With the passive sign convention implicit in that figure 
(we use this convention to treat impedances and sources
in a uniform manner)
we get
$v=zi+v_s$, that is,
\begin{equation}
v-zi =v_s.\label{vsource}
\end{equation}
With $z=0$ this amounts to $v=v_s$ and we get an ideal voltage source.

The dual case (Fig.\ 1(b))
is defined by a current source with an 
impedance in parallel, typically represented by its admittance parameter
$y$. Here the relation reads as
\begin{equation}
yv - i=i_s, \label{csource}
\end{equation}
with the value $y=0$ modelling an ideal current source.
Note that in both cases 
we want to encapsulate both the source and the impedance (if
present) in a single circuit element. 

Both (\ref{vsource}) and (\ref{csource}) are particular
instances of a general equation of the form
\begin{equation}
pv-qi=s \label{affine}
\end{equation}
with $p,$ $q,$ $s \in \K$ and the restriction that $p$, $q$ do
not vanish simultaneously. In particular, provided that $p$ 
does not vanish one gets (\ref{vsource}) 
from (\ref{affine}) by setting
$z=q/p$, $v_s=s/p$ (analogously, if $q \neq 0$ we get (\ref{csource}) with
$y=p/q$, $i_s=s/q$). If neither $p$ nor
$q$ do vanish, then both descriptions
(\ref{vsource}) and (\ref{csource}) are possible and the well-known
Th\'evenin-Norton identities $v_s=zi_s$, $z=1/y$ apply. By contrast,
the cases $q=z=0$ and $p=y=0$ describe ideal sources and only 
(\ref{vsource}) or (\ref{csource}), respectively, do hold,
whereas (\ref{affine}) accommodates both. 
It is worth mentioning
that the formalism above makes also sense (in the real case) 
for time-dependent sources $s(t)$.

A key aspect in our approach is that (\ref{affine}) 
is 
again defined up to a non-vanishing
multiplicative factor, because the zero set
of  $pv-qi-s$ does not change
if we multiply this polynomial (in $i$, $v$) 
by a non-zero constant. This makes it natural to set
now $V$ as the (three-dimensional) vector space of 
polynomials in two indeterminates with coefficients in $\K$
and degree not greater than one, and assume that 
the characteristic 
lies on 
the projective space $\PP (V)$ resulting from the equivalence
relation (\ref{equiv}).
Additionally, if we denote by 
$[k]$ 
the equivalence class
defined by all non-zero constants in $\K$, 
the requirement that $p$ and $q$ do not vanish simultaneously means that the
equivalence classes we are interested in belong 
to the punctured space $\PP(V)-\{[k] \}$.
This is the set of equivalence classes 
corresponding to polynomials of degree (exactly) one.
Moreover, it is clear that in the basis defined by the polynomials
$f_1(i, v)=v$, $f_2(i, v)=-i$, $f_3(i, v)=-1$,  
the homogeneous coordinates 
are
$(p:q:s)$, with $(p,q) \neq (0,0)$ by the aforementioned requirement.


Finally, we identify $\PP(V)$ with the projective 
plane $\K\PP^2$ by means of the isomorphism $V \to \K^3$
defined by
$\alpha v + \beta i + \gamma \to (\alpha, -\beta, -\gamma)$
(and note that, again, the homogeneous coordinates above correspond to the ones 
obtained in $\K\PP^2$ from the choice of the canonical basis in $\K^3$).
With this isomorphism in mind, we will think of sources as points lying
on the projective plane $\K\PP^2$.
More precisely, because of the non-vanishing requirement on 
(at least) either $p$ or $q$,
our description of abstract sources is finally bound to
lie on the punctured projective plane 
$\K\PP^2_*=\K\PP^2-\{(0:0:1)\}$.

\subsection{Abstract linear elements}
\label{subsec-abstract}

It is not by chance that the discussion in subsection \ref{subsec-sources}
generalizes the one
in \ifthenelse{\boolean{ieee}}{}{subsection }\ref{subsec-impedances}. Indeed, the
relation defined 
by (\ref{affine}) comprises all possible cases and
sets up an 
abstract linear element 
whose electrical nature is not determined a priori;
it is fixed only after 
choosing values for the homogeneous parameters
$p$, $q$ and $s$. Two natural taxonomies arise. 

The first taxonomy depends on the parameter $s$; the device amounts to 
a source when an assignment $s \neq 0$ is made, or to a
linear impedance 
when $s=0$. 
In the first case ($s \neq 0$), it is not necessary to
recast the source as a voltage source or as a current source: actually,
it is important to understand that (\ref{affine}) may define
an abstract source of neither type, even if locally (that is,
for specific values of $p$, $q$) it can always be reduced either
to a voltage-source form (by setting $p=1$)
or to a current-source form ($q=1$); if both parameters are non-zero
then both forms are admissible.
But there is no need a priori to do this reduction.
Still in the $s \neq 0$ patch 
we may further classify a source as an ideal (resp.\ a non-ideal) source
if either $p$ or $q$ vanishes (resp.\ none of them vanishes).
When $s=0$ the circuit element behaves as an impedance and, again,
there is no need to specify if it 
is current- or voltage-controlled: 
both options
exist except in the, say, extremal cases $p=0$ (open-circuit) and $q=0$ 
(short-circuit). 

The second taxonomy classifies linear elements
according to the
values of $p$, $q$, disregarding $s$. The patch $\mathbb{A}_z$,
defined by the condition
$p \neq 0$,
accommodates the set of current-controlled elements, including in particular
short-circuits and also
voltage sources (more precisely, abstract sources that admit a description
as voltage sources, either non-ideal or ideal).
Assignments
with $q \neq 0$ (defining the patch $\mathbb{A}_y$) correspond
to voltage-controlled elements, including open-circuits
and sources that admit a description as current sources.




\subsection{Linear circuits as digraphs with
projective weights. Configurations}
\label{subsec-config}

In order to lift the projective formalism of previous subsections
to the circuit level
we only need
to combine this approach with the classical description of circuits
in terms of directed graphs. The reader is referred to 
\cite{bollobas, chen, chenFilters, 
wsbook} for background
in this regard.


A linear circuit 
is simply modelled in our framework as a 
directed
graph $G$ endowed with a map $\gamma: E(G) \to \K\PP^2_*$, that
is, a map assigning
to each branch $e \in E(G)$ a point in the 
punctured
projective plane $\K\PP^2_*$ 
defined above. In terms of homogeneous coordinates, we assign
to every branch $e \in E(G)$ a triad of (homogeneous)
parameters $(p:q:s)=\gamma(e)$, with $(p,q) \neq (0,0)$. 
We call $\gamma$ a {\em configuration
map}. 

A 
{\em source-free configuration} is a map $\hspace{0.5mm}\check{\hspace{-0.5mm}\gamma}: E(G) \to \K\PP$
assigning to every branch $e$ a projective point
with homogeneous coordinates $(p:q)=\hspace{0.5mm}\check{\hspace{-0.5mm}\gamma}(e)$. 
Via the  projection map
\begin{eqnarray} \label{projection} 
\pi: && 
\ifthenelse{\boolean{ieee}}{\K \PP^2_* \hspace{3.4mm} \to \hspace{3mm} \K\PP \\ \nonumber
&& \hspace{-5mm} (p:q:s) \hspace{1.5mm}  \to \hspace{1mm} (p:q),}
{\K \PP^2_* \hspace{3.4mm} \to \hspace{3mm} \K\PP \\ \nonumber
&& \hspace{-5mm} (p:q:s) \to (p:q),}
\end{eqnarray}
we associate
to any 
configuration $\gamma$ a source-free
configuration via 
the 
relation
$\hspace{0.5mm}\check{\hspace{-0.5mm}\gamma}=\pi \circ \gamma$. In practice,
this amounts to setting
$s=0$ for all branches 
or, in classical terms, 
to replacing voltage (resp.\ current) sources
by short (resp.\ open) circuits.
Note that (\ref{projection}) is well-defined 
because $(0:0:s) \not\in \K\PP^2_*$.

We may further assume that the set of branches $E(G)$ just amounts 
to $\{1, \ldots, m\}$. This way, the whole set of parameter
values is described as a point in the space
$\K\PP^2_* \times \hspace{1mm}\stackrel{(m)}{\ldots}\hspace{1mm} \times \K\PP^2_*$. We will call this Cartesian product
the {\em configuration space} or the {\em parameter space} and,
for simplicity in the notation, from now on
we skip the symbol $m$ in such products.
In the absence of sources, the parameter set takes values 
in the {\em source-free configuration space}
$\K\PP \times \ldots \times \K\PP$. 
This is often called a {\em multiprojective} space and, as a cautionary
remark, the
reader should be aware of the fact that it is 
not equivalent (if $m > 1$) to $\K\PP^m$ (e.g.\ 
$\K\PP \times \K\PP$ is {\em not}
the projective plane $\K\PP^2$). 







\section{Homogeneous circuit models}
\label{sec-homog}



\begin{versionC}
\color{red}
See Chen Eng. Appls. pp 59 y ss. 

{\bf P 59: generalized Ohm's law} (llevar arriba), but only $v=v_s + Ri$ or the
other; ie assumes a priori a voltage-controlled form or vv (2.42a OR 2.42b). 
Source transformations.

Mencionar arriba tb fuentes $x(t)$

And then {\em branch-current system of equations}:
\begin{eqnarray}
A i & = & 0\\
B Z i &= & -BE
\end{eqnarray}
y dice es el que usaba Kirchhoff en origen.
Dual: {\em branch-voltage system of equations}:
\begin{eqnarray}
A Yv & = & -AJ\\
B v &= & 0
\end{eqnarray}

Maxwell: usaba {\em loop system of equations}:
$$BZB^{\tra} j = -BE$$

{\em Cut system of equations,} ``cut-voltage vector'' etc... Node-pair voltages, Kron... Q becomes A: node to datum voltages etc.
$$QYQ^{\tra} e = -QJ \text{ (rev.) }$$

\

\color{black}
\end{versionC}














The formalism presented in Section \ref{sec-proj} makes it  
possible to introduce a comprehensive
framework where different circuit model families 
can be properly placed and analyzed. 
We undertake this task in the present section, where
in particular we formulate an $m$-dimensional reduction of linear
circuits 
holding without restrictions (cf.\ (\ref{homogsym})).

\subsection{
General homogeneous model. 
Characteristic space}
\label{subsec-gen}


We refer the reader to \cite{bollobas, chen, chenFilters, chuadesoerkuh,
wsbook}
for background on the topics here discussed. 
We will assume
for simplicity that the digraphs/circuits involved in all results 
are connected. Denote by $n$ and $m$ the number
of nodes and the number of branches in the circuit.

We assume that the reader is familiar with the full
incidence matrix $\mathbf{A} \in \R^{n \times m}$ of a directed graph, 
with entries
defined as $a_{ij}=+1$ (resp.\ $-1$) if the $j$-th branch leaves
(resp.\ enters) the $i$-th node, and $0$ otherwise. Denote by
$\mathbb{I}=
\ker \mathbf{A} \subseteq \R^m$ 
the so-called {\em cycle space}, which is known
to comprise all current vectors solving Kirchhoff's current
law. Analogously, the {\em cut space}
$\mathbb{V}= \im \mathbf{A}\hspace{-0.5mm}^{\tra} 
\subseteq \R^m$
defines the set of voltage vectors solving Kirchhoff's voltage law.
Both spaces are orthogonal to each other, 
according to Tellegen's identity. In a connected digraph,
we have $\dim \mathbb{I} = m-n+1$,  $\dim \mathbb{V} = n-1$. 

Now, let $A \in \R^{(n-1) \times m}$ and $B \in \R^{(m-n+1)\times m}$ 
be maximal rank matrices with entries 
in $\{\pm 1, 0\}$ 
which satisfy
$\ker A = \mathbb{I}$, $\ker B = \mathbb{V}$,
so that Kirchhoff laws can be written as the independent
sets of linear equations $Ai=0$, $Bv=0$.
We call $A$ a 
{\em cut matrix} and $B$ a {\em cycle matrix}.
Typically, $A$ is 
a reduced incidence matrix or a reduced cutset matrix of the 
digraph and 
$B$ can be chosen as a reduced 
loop matrix.


In order to join together 
the different linear elements defining the circuit,
we assume that $(p_k:q_k:s_k)$ are homogeneous coordinates
describing the $k$-th circuit element and
let $P$ and $Q$ be diagonal matrices of order $m$ with 
$p_k$ and $q_k$, respectively, in the $k$-th diagonal
position. We write as $s$ the vector defined by the
excitation terms $(s_1, \ldots, s_m)$
(set also 
$p=(p_1, \ldots, p_m)$ and
$q=(q_1, \ldots, q_m)$).
Recall that all these parameters may take values either in $\R$
(for resistive circuits with DC sources) or in $\C$ (AC circuits
in sinusoidal steady state, or circuits modelled in the Laplace domain),
and that 
$\K$ 
is either $\R$ or $\C$. 
In the light of subsection \ref{subsec-config},
the set of homogeneous parameters $((p_1:q_1:s_1), \ \ldots, \ (p_m:q_m:s_m))$
defines a point in the multiprojective space 
$\K\PP^2_* \times \ldots \times \K\PP^2_*$.

The fact that $p_k$ and $q_k$ cannot vanish simultaneously
is reformulated here by requiring
the $m \times 2m$ matrix $\begin{pmatrix}P & Q\end{pmatrix}$
to have maximal rank.
For later use, the fact that each homogeneous triad
$(p_k: q_k : s_k)$ is defined
up to a non-null constant (say $d_k  \in \K -\{0\}$)
is recast 
by saying that $P$, $Q$, $s$ are defined
up to premultiplication by a non-singular, diagonal, order $m$ matrix $D$, 
with 
$d_k$ in the $k$-th diagonal position. 
Be also aware
of the fact that diagonal matrices commute, so that $DP=PD$, $DQ=QD$
and also $PQ=QP$.

In these terms,
the equations
of an arbitrary (uncoupled, time-invariant) linear circuit 
can be 
written as
\begin{equation} \label{generalbranch}
\begin{pmatrix} A & 0 \\ 0 & B \\ -Q & P \end{pmatrix}
\begin{pmatrix} i \\ v \end{pmatrix}
=
\begin{pmatrix} 0 \\ 0 \\ s \end{pmatrix}.
\end{equation}



Let us define $\mathbb{M}$ as the 
$m$-dimensional 
affine subspace of $\K^{2m}$ defined by the set of current-voltage pairs which
satisfy the characteristic equations of all circuit branches, that
is, 
\begin{equation}
\mathbb{M} = \{ (i,v) \in \K^{2m} \ / \ -Qi+Pv=s\}.
\end{equation}
We will call
this the {\em characteristic space}. 
Clearly, the solutions of the circuit equations (\ref{generalbranch})
are the current-voltage vectors which belong to the 
intersection space
$(\mathbb{I} \times \mathbb{V}) \cap \mathbb{M}$.

The circuit equations (\ref{generalbranch}) have $2m$ unknowns
and an order reduction is advisable.
Many (if not all) families of circuit analysis models may be 
understood as a reduction of the circuit equations 
(\ref{generalbranch}) to one of the spaces above (that is,
the ``Kirchhoff space'' $\mathbb{I} \times \mathbb{V}$ or the characteristic
space $\mathbb{M}$), or to
a certain subspace of one of them, after a suitable parametrization is given.
Traditionally, in the literature such reductions are performed
under certain assumptions on the controlling variables (an example of 
such an assumption would
be ``all devices are voltage-controlled'', what in a classical
framework means that voltage sources are excluded and that 
impedances have an admittance description, to get a reduction
in terms of voltages; precise details are given later). 
Our goal is either to perform such reductions in total generality,
something that will be achieved by an appropriate reduction to
the characteristic space $\mathbb{M}$ or, in other cases,
to give a precise characterization of
the configurations allowing for whatever reduction. This framework 
can be also used in the characterization of nodal- and loop-analysis models.

\subsection{Classical 
reductions}
\label{subsec-classical}

Focus on the subspace of the multiprojective configuration space
$\K\PP^2_* \times \ldots \times \K\PP^2_*$ defined by
the condition that all $p_k$ coordinates are non-null or, equivalently,
that the matrix $P$ defined above is non-singular (this
defines the {\em current-controlled patch} $\mathbb{A}_{\mathbf{z}}$;
we
use the boldface symbol $\mathbf{z}$ to distinguish it from the corresponding
``one-element'' patch
$\mathbb{A}_z \subseteq \K\PP^2_*$, with $\mathbb{A}_{\mathbf{z}} =
\mathbb{A}_z \times \ldots \times \mathbb{A}_z$).
In this patch, all sources admit a 
classical
description as voltage sources,
with excitation vector $v_s=P^{-1}s \in  \K^m$; 
branches for which $s_k=0$,
which correspond to impedances,
contribute a null entry to the vector $v_s$. Additionally,
the impedance parameter of all branches, including sources, 
is well-defined and can be described by means
of the impedance matrix $Z=P^{-1}Q$: this applies in particular to
non-ideal source branches (in these cases $z_k$ is the series impedance
of Fig.\ 1(a)) and also to ideal voltage sources (for which $z_k=0$).
It is easy to check that both
\begin{equation} \label{Zvs}
Z=P^{-1}Q, \ v_s=P^{-1}s
\end{equation} 
are independent of the choice of homogeneous
coordinates.

In essence, what we are using is
a parametrization of the characteristic space
$\mathbb{M}$ 
by the current variable $i$, namely through the relation 
\begin{equation}
v=Zi+v_s, \label{curcol}
\end{equation}
with $Z$ and $v_s$ given by (\ref{Zvs}).
This yields the so-called {\em branch-current model}
\begin{subequations}\label{branchcurrent}
\begin{eqnarray} 
Ai & = & 0 \\
BZi & = & -Bv_s.
\end{eqnarray}
\end{subequations}
The key idea is that both the parametrization (\ref{curcol}) and
the branch-current model (\ref{branchcurrent}) are well-defined
only for parameter sets lying on the patch $\mathbb{A}_{\mathbf{z}}$.
Incidentally, splitting the matrices $A$, $B$ by columns (according
to the source/impedance nature of branches), and redefining
$v_s$ accordingly, one gets the classical form of the branch-current model,
cf.\ \cite{chen}. But in our formalism we do not need
to make this splitting; the entries of $Z$ and $v_s$ implicitly perform
this task.

In the dual case, under the assumption that 
$Q$ is non-singular, so that
the admittance matrix $Y=PQ^{-1}$ is well defined
and sources admit a description as current sources with $i_s=Q^{-1}s$, 
one gets
the {\em branch-voltage model} (cf.\ again \cite{chen}
in the classical setting)
\begin{subequations} \label{branchvoltage}
\begin{eqnarray} 
AYv & = & Ai_s \\
Bv & = & 0.
\end{eqnarray}
\end{subequations}
Again, this can be seen as a reduction of the circuit
model to the characteristic space $\mathbb{M}$,
in this case by means of the parametrization $i=Yv-i_s$.
This reduction is only valid on the voltage-controlled patch $\mathbb{A}_{\mathbf{y}}
\subseteq \K\PP^2_* \times \ldots \times \K\PP^2_*$.

Finally, even if we omit details for the sake of brevity,
hybrid models (see \cite{GR14} in this regard),
which combine current- and voltage-controlled descriptions,
can be also described as a reduction of (\ref{generalbranch})
under appropriate parametrizations of the characteristic space $\mathbb{M}$.
But also in this case 
an {\em a priori} assignment of a control variable to all branches
is still necessary, so that neither these models can be used on the 
whole configuration space $\K\PP^2_* \times \ldots \times \K\PP^2_*$.

\subsection{Homogeneous reductions}
\label{subsec-homog}

Contrary to the 
models above, the homogeneous 
formalism makes it possible
to perform an $m$-dimensional reduction without any working assumption
on controlling variables, and accommodating
all possible cases (even degenerate  ones; cf.\ Section \ref{sec-nondeg}),
that is, applying on the whole configuration space
$\K\PP^2_* \times \ldots \times \K\PP^2_*$.
To do so we introduce 
abstract variables $u \in \K^m$, 
to be termed 
{\em
homogeneous 
  variables} (cf.\ subsection \ref{subsec-u} below),
which parametrize the characteristic space $\mathbb{M}$
in the form
\begin{subequations} \label{branchparam}
\begin{eqnarray} 
i & = & Pu + i_0 \label{branchparami}\\
v & = & Qu + v_0. \label{branchparamv}
\end{eqnarray}
\end{subequations}
For these relations to actually parametrize $\mathbb{M}$,
the vector $(i_0, v_0)$ needs to solve the equation $-Qi+Pv=s$.
This means that $(i_0, v_0)$ is itself a 
point 
of $\mathbb{M}$. 
We will call it the {\em origin} of the parametrization
(the terminology just reflects that (\ref{branchparam}) sets up an affine
system of coordinates for $\mathbb{M}$) and 
will elaborate on possible choices later on.

By inserting (\ref{branchparam}) into (\ref{generalbranch}) 
we get a general {\em homogeneous branch model} of the form
\begin{eqnarray}  \label{homoggen}
\begin{pmatrix}AP \\ BQ\end{pmatrix}u=
\begin{pmatrix}-Ai_0 \\ -Bv_0\end{pmatrix},
\end{eqnarray}
which is defined without the need to impose any restriction on $P$, $Q$.
Currents and voltages are simply obtained from the solutions $u$ of 
(\ref{homoggen}) as $i =  Pu + i_0$, $v = Qu + v_0$ (cf.\ (\ref{branchparam})).

\begin{versionC}
\color{red}
THIS IS OLD. SHOULD BE $v_0=v_s$ ETC. WORKS WELL, 28.3.18(2)

Particular choices of $P$, $Q$, $i_0$ and $v_0$, holding under certain
assumptions, yield the  branch-voltage and branch-current models
above. Current-controlled setting: $P=I$, $i_0=0$, $v_0=s$ should 
yield the branch-current model... [BUT THE LATER
ONLY BECAUSE $P=I$, MAKES IT CONFUSING BECAUSE THERE IS NO $P^{-1}$ IN
THE MODEL ARISING HERE]. VER COMO PLANTEO ESTO.

A way to do it is to reformulate the branch-model with $Du$ instead
of $u$ in the parametrization, but this makes it more complicated

**OLDER: Note that other choices for the right inverse $F^+$ of $\begin{pmatrix}-Q & P\end{pmatrix}$
(with the notation...)
are possible. In particular, assume that $Q$ is invertible: we may then
choose the right inverse in the form 
$\begin{pmatrix}-Q^{-1} & 0\end{pmatrix}^{\tra}$ and this leads to
\begin{subequations} \label{characred2}
\begin{eqnarray} 
i & = & Pu -Q^{-1}s \label{characred2i}\\
v & = & Qu \label{characred2v}
\end{eqnarray}
\end{subequations}
with the reduction reading as
\begin{subequations} \label{characequ2}
\begin{eqnarray} 
APu & = & AQ^{-1}s \label{characequ2a}\\
BQu & = & 0. \label{characequ2b}
\end{eqnarray}
\end{subequations}
Even exactly the branch-voltage? Should be via another choice of the homogeneous coordinates,
namely $\begin{pmatrix}PQ^{-1} & I \end{pmatrix}$. Relate with the aforementioned fact that
the characteristic variables are defined only up to a diagonal isomorphism (``(anisotropic) scaling'')... define $D$ (scaling matrix)

Let us compare, with the general choice (\ref{characred}), the characteristic equations
(\ref{characequ}) with the branch-voltage equations (\ref{branchvoltage}),
which are defined only if $Q$ is non-singular: in this setting $u$ is
not the voltage $v$ but (\ref{characred}) is a linear 
change of coordinates mapping $u$ onto $v$ and transforming 
(\ref{branchvoltage})
into (\ref{characequ}) --check: from
$AYv =  AQ^{-1}s$ and using $Y=PQ^{-1}$ [probablemente m\'as
f\'acil al rev\'es] write
\begin{eqnarray*}
AY(Qu + P(P^2+Q^2)^{-1}s)) & = & APu+AQ^{-1}P^2(P^2 + Q^2)^{-1}s\\
& = & APu+AQ^{-1}(P^2+Q^2-Q^2)(P^2 + Q^2)^{-1}s\\
& = & APu+AQ^{-1}s - AQ(P^2 + Q^2)^{-1}s
\end{eqnarray*}
which equals $AQ^{-1}s$ yielding (\ref{characequa}).**
\color{black}
\end{versionC}

\begin{versionC}
\color{red}
\subsection{On the $u$-variables}

Avoid thinking it of either the voltage or the current... In general, a 
linear (of affine) combination of both which yields a well-defined 
reduction {\em for all configurations} (by contrast to $v$ or $i$
in the branch-voltage and branch-current models, which require $Q$ and $P$
to be non-singular, respectively). Specifically, from (\ref{branchparam})
one easily gets
\begin{eqnarray} \label{defu}
u = (P^2+Q^2)^{-1}\begin{pmatrix} P & Q\end{pmatrix}
\begin{pmatrix}i-i_0 \\ v-v_0\end{pmatrix}
\end{eqnarray}
(or write it as $P(i-i_0)+Q...$)
\color{black}
\end{versionC}

The models of subsection \ref{subsec-classical} can be derived 
as particular instances of 
(\ref{homoggen}). Focus for example on the branch-current model
\eqref{branchcurrent}, which holds under the assumption that $P$
is non-singular. The latter means that the choice $\hat{P}=I$ is 
admissible for the homogeneous description of 
circuit elements; the remaining parameters then read as $\hat{Q}=Z$, 
$\hat{s}=v_s$ (cf.\ \eqref{Zvs} and \eqref{curcol}). 
Together with 
$\hat{i_0}=0$, $\hat{v_0}=v_s$,
one can easily check that \eqref{branchparam}
reads as $i=u$, $v=Zu+v_s$ and that \eqref{homoggen} amounts
to the branch-current model (\ref{branchcurrent}). Similar
remarks apply to 
the branch-voltage
model (\ref{branchvoltage}), 
which can be derived analogously from (\ref{homoggen}) under appropriate
assumptions. 

\subsection{Symmetric form} 



The origin $(i_0, v_0)$ can be always chosen, without
any further assumptions on $P$, $Q$, as
\begin{subequations} \label{symbase}
\begin{eqnarray} 
i_0 & = & -Q(P^2+Q^2)^{-1}s \label{symbasei}\\
v_0 & = & P(P^2+Q^2)^{-1}s, \label{symbasev}
\end{eqnarray}
\end{subequations}
where we use the fact that the (diagonal) matrix
$P^2 + Q^2$ is invertible because 
$(p_k, q_k) \neq (0, 0)$ for all circuit branches.
%
With the choice (\ref{symbase}), the homogeneous branch model
takes the (say) symmetric form
\begin{eqnarray} \label{homogsym} 
\begin{pmatrix} AP \\ BQ \end{pmatrix}
u = \begin{pmatrix} AQ \\ -BP \end{pmatrix}
(P^2+Q^2)^{-1}s.
\end{eqnarray}

We
will sometimes write $\bar{s}=(P^2+Q^2)^{-1}s$ 
for notational simplicity. 
We emphasize that this reduced model is well-defined for any linear
circuit, even for degenerate ones; that is, no restriction in the 
the circuit parameters is necessary for the model
(\ref{homogsym})
to hold. This makes it useful for analytical purposes and
suitable for computational implementation.
Note that
the inverse involved is that of a diagonal matrix  and hence poses no
computational difficulties. 
The branch currents and voltages
are recovered from the solutions of (\ref{homogsym}) from
the relations depicted in
(\ref{branchparam}) and (\ref{symbase}), which altogether yield
\begin{eqnarray} \label{branchparamsym}
\begin{pmatrix}i \\ v\end{pmatrix}=
\begin{pmatrix}P \\ Q\end{pmatrix}u
+ \begin{pmatrix}-Q \\ P\end{pmatrix}(P^2+Q^2)^{-1}s.
\end{eqnarray}
%






\subsection{
Homogeneous variables}
\label{subsec-u}

Equations \eqref{homoggen} and \eqref{homogsym} provide 
completely general models for an $m$-dimensional description 
of linear circuits, not requiring any assumptions
on the existence of current-controlled or voltage-controlled 
descriptions of the circuit elements. 
The price, of course, is the lack of a 
physical meaning on the unknowns, namely, the $u$ variables, in contrast to
the current $i$ and the voltage $v$; the variables $u$ can be
thought of as a ``seed'' from which both the current and the
voltage stem by means of the relations \eqref{branchparam} or 
\eqref{branchparamsym}. More prosaically, we will say that $u$ is
a vector of {\em homogeneous variables},
borrowing the term from the homogeneous nature of the
parameters $P$, $Q$, and $s$. There is a terminological abuse here;
see however the remarks in the following paragraph. 

The 
variables 
$u$ are not uniquely defined,
because they arise as the unknowns in \eqref{homoggen} or 
\eqref{homogsym} and the actual form of both systems 
(in other words, 
the 
coordinate system parametrizing $\mathbb{M}$)
depends on the choice of 
$P$, $Q$, $s$ 
and 
$(i_0, v_0)$.
There is no problem with this because
$u$ is always accompanied by
$P$, $Q$ when 
recovering 
$i$, $v$ via (\ref{branchparam})
or (\ref{branchparamsym}): 
in practice,
the simpler choice of $P$,
$Q$, the better.
In any case, 
two different choices of these parameters
(say $P$, $Q$, $i_0$, $v_0$ and
$\tilde{P}$, $\tilde{Q}$, $\tilde{i_0}$, $\tilde{v_0}$)
yield two sets of homogeneous variables which result
from a diagonal rescaling of one another. Indeed,  
the identities $\tilde{P}=PD$, $\tilde{Q}=QD$ must hold
for some 
scaling matrix $D$, as indicated
in subsection \ref{subsec-gen}, and one can show that this implies that
under the symmetric choice (\ref{branchparamsym})
the relation
$u=D\tilde{u}$
holds. This means that each scalar variable
$u_k$ is defined up to a non-vanishing constant, exactly
as the 
parameters $p_k,$ $q_k,$ $s_k$ are: this 
(informally) gives the $u$ variables a homogeneous flavor.
For the sake of completeness, be aware that in general
(more precisely,
when the choice of parameters yields a difference 
vector 
$(\tilde{i_0}-i_0, \tilde{v_0}-v_0)$
not belonging to 
$\ker \begin{pmatrix}P& Q\end{pmatrix}$) 
the equation relating $u$ and $\tilde{u}$ can be proved 
to have the form
\begin{eqnarray} \label{branchbis2}
u = D \tilde{u} + (P^2+Q^2)^{-1}
\begin{pmatrix}P & Q\end{pmatrix}
\begin{pmatrix}\tilde{i_0}-i_0 \\ \tilde{v_0}-v_0\end{pmatrix}.
\end{eqnarray}

\subsection{Elementary examples. Partially homogeneous models}
\label{subsec-ex}

We illustrate the ideas above by means of a very simple
example, depicted in Fig.\ 2(a).

\begin{figure}[ht] \label{fig-series}
\ifthenelse{\boolean{ieee}}{\hspace{-5mm}}{\hspace{15mm}} 
\parbox{0.5in}{
\ifthenelse{\boolean{ieee}}{\epsfig{figure=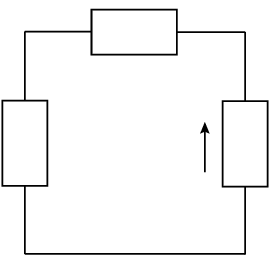, width=0.20\textwidth}}
{\vspace{6mm}\epsfig{figure=3cycle.eps, width=0.234\textwidth}}
\ifthenelse{\boolean{ieee}}{
\put(-96,40){\small $1$}
\put(-54,83){\small $2$}
\put(-12,40){\small $3$}
}
{\put(-103,44){\small $1$}
\put(-59,88.4){\small $2$}
\put(-13,44){\small $3$}
}
}
\ifthenelse{\boolean{ieee}}{\hspace{27mm}}{\hspace{55mm}}
\parbox{0.5in}{\ifthenelse{\boolean{ieee}}{\vspace{0.12mm}}{\vspace{5mm}}
\ifthenelse{\boolean{ieee}}{\epsfig{figure=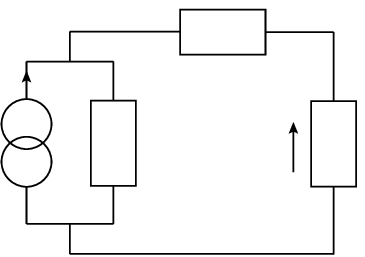, width=0.2658\textwidth}}{\epsfig{figure=series_bis.eps, width=0.315\textwidth}}
\ifthenelse{\boolean{ieee}}{
\put(-97,41){\small $y_1$}
\put(-54.5,82.5){\small $2$}
\put(-12.5,40.5){\small $3$}
}
{
\put(-106.3,46){\small $y_1$}
\put(-60,90){\small $2$}
\put(-14,45){\small $3$}
}
}
\ifthenelse{\boolean{ieee}}{\caption{\hspace{-1mm}(a) A 3-loop. \hspace{6mm} (b) 
Assume the first branch to be a current source.}}{\caption{\hspace{6mm}(a) A 3-loop. \hspace{16mm} (b) Assume the first branch to be a current source.\hspace{13mm}}}
\end{figure}

\vspace{-1mm}

We show here how to set up the models by hand (whereas
the examples in Section \ref{sec-faults} make systematic use of the
digraph matrices $A$, $B$). Direct all branches 
counterclockwise, and write Kirchhoff laws
as $i_1 = i_2= i_3$ and $v_1 +  v_2 +  v_3=0$. 
We assume that branches 2 and 3 do not accommodate sources (i.e.\ $s_2=s_3=0$)
but that branch 1 does. 
The relations (\ref{branchparamsym}) read as

\ifthenelse{\boolean{ieee}}{\vspace{-5mm}}{\vspace{-7mm}}
\begin{subequations}
\begin{eqnarray}
i_1 & = & p_1 u_1 - \frac{q_1}{p_1^2 + q_1^2}s_1 \\
v_1 & = & q_1 u_1 + \frac{p_1}{p_1^2 + q_1^2}s_1
\end{eqnarray}
\end{subequations}
for the first branch, and $i_k = p_k u_k$, $v_k = q_k u_k$ for $k=2,$ $3$. 
Inserting all of them in Kirchhoff's
relations we get the symmetric model (\ref{homogsym}), which here
has the form
\begin{subequations}
\begin{eqnarray}
p_1 u_1 - p_2 u_2 & = & \frac{q_1}{p_1^2 + q_1^2}s_1 \\
p_2 u_2 - p_3 u_3 & = & 0 \\
q_1 u_1 + q_2 u_2 + q_3 u_3 & = &  - \frac{p_1}{p_1^2 + q_1^2}s_1.
\end{eqnarray}
\end{subequations}
The determinant of the coefficient matrix (which corresponds to the one on the left-hand-side
of (\ref{homogsym})) is 
\begin{equation} \label{polyn1}
p_1 p_2 q_3 + p_1 q_2 p_3 + q_1 p_2 p_3, 
\end{equation}
which already arose
in the Introduction (find a detailed discussion on the
form of this polynomial 
in Section \ref{sec-nondeg}). Note that for the moment 
we are not assuming either
a voltage- or a current-controlled form for any device, hence the
multihomogeneous form of the polynomial above.

In most practical cases there would be no loss of generality in 
giving the source 
a classical form, e.g.\ as a (non-ideal) current source, as 
in Fig.\ 2(b). 
In this setting there is no advantage in keeping a
homogeneous variable $u_1$ in the model; it
is more convenient to describe the source simply by means of the 
relation $i_1=y_1 v_1 -i_s$ 
and
use $v_1$ as a model variable. Retaining the homogeneous form for the other two branches 
(to include simultaneously all possible cases, including short- and open-circuits) we
get a partially homogeneous model
\begin{subequations}
\begin{eqnarray}
y_1 v_1 - p_2 u_2 & = & i_s \\
p_2 u_2 - p_3 u_3 & = & 0 \\
v_1 + q_2 u_2 + q_3 u_3 & = & 0,
\end{eqnarray}
\end{subequations}
with unknowns $(v_1, u_2, u_3)$.


\begin{versionC}
\color{red}
In greater generality (rev.): ``classical'' form for sources (subscript j current
sources, v voltage sources)
$i_j=Y_jv_j -i_s$, $v_v=Z_v i_v + v_s$
(ideal or not: $y_j=0$ for ideal current sources, $z_v=0$ for ideal
voltage sources), ``homogeneous form'' for
impedances (subscript h). Denote $u$ for homogeneous variables at the 
impedance branches
\begin{eqnarray}
A_h P_h u + A_j Y_jv_j + A_v i_v & = & A_j i_s \\
B_h Q_h u + B_j v_j + B_v Z_v i_v & = & - B_v v_s.
\end{eqnarray}
\color{black}
\end{versionC}

Partially homogeneous models can be also used in the context of
other circuit analysis techniques, as illustrated
in the sequel. Assume for instance that all impedances
in Fig.\ 2(b) are known to have an admittance description
(i.e.\ we work in the patch $\mathbb{A}_{\check{\mathbf{y}}}$), and
that we want to compute 
the Th\'evenin equivalent across the third branch. 
Our goal is 
to set up only one model allowing us to compute both the
Th\'evenin (open-circuit) voltage and the Norton (short-circuit) current.
To this end, we simply model
a virtual load in parallel with the third branch 
in homogeneous terms, 
and use a classical description for the remaining branches.
The nodal analysis model can be checked to read as
\begin{subequations} \label{MNA}
\begin{eqnarray}
y_1 e_1 + y_2 (e_1-e_2) & = & i_s \\
y_2 (e_2-e_1) + y_3 e_2 - p_lu_l & = & 0 \\
e_2  + q_l u_l & = & 0,
\end{eqnarray}
\end{subequations}
with unknowns $(e_1, e_2, u_l)$;
$e_1$ and $e_2$ are the potentials at the NW and NE nodes and
the subscript $l$ is used for the (virtual) load. Solving this system
with $p_l=0$ yields the Th\'evenin voltage as $q_lu_l$,
whereas the Norton (short-circuit) current is obtained as
$p_lu_l$ with $q_l=0$. 

More conclusions can be derived from the latter model.
The determinant is now $(y_1 y_2 + y_1 y_3 + y_2 y_3)q_l+ (y_1 + y_2)p_l$.
When $p_l=0$ (so that $q_l \neq 0$), the non-vanishing of the first term
requires $y_1 y_2 + y_1 y_3 + y_2 y_3 \neq 0$, which characterizes the
set of 
configurations in $\mathbb{A}_{\check{\mathbf{y}}}$ for which the Th\'evenin voltage (and the
Th\'evenin equivalent circuit) is well-defined; analogously, 
when $q_l=0$ (and then $p_l \neq 0$) the non-vanishing of the
above polynomial requires $y_1 + y_2 \neq 0$, a condition which characterizes
the parameter values (again in $\mathbb{A}_{\check{\mathbf{y}}}$) for which the Norton current (and the Norton equivalent)
are well-defined.




\section{Non-degeneracy. The matrix-tree theorem}
\label{sec-nondeg}

The homogeneous formalism makes it possible to address in 
general
the non-degeneracy problem, that is, the characterization
of the conditions under which a 
linear circuit has a unique
solution. This is e.g.\ of interest in non-passive problems, in which 
the real part of some impedances 
(or the resistance in a DC context)
may become negative.
These properties have been typically examined in circuit theory in terms of
certain reduced models which, as discussed earlier, are not completely general.
In particular, in nodal or loop analysis the results are formulated in terms
of the nodal admittance matrix or the loop-impedance matrix, respectively
\cite{chen}. 
In the nodal or the loop analysis 
context, the degeneracy of a circuit
is characterized by 
the zeros of a tree- or a cotree-enumerator polynomial,
respectively. However, such polynomials provide no explicit information about what happens
when the assumptions supporting such reductions do not hold. 

In this section we perform this analysis in full generality,
using a projective version of the aforementioned matrices
(cf.\ (\ref{APBQ})) and the
 multihomogeneous
version (\ref{kir-hom}) of the Kirchhoff or tree-enumerator polynomial:
find related results regarding the latter
in a matroid context in \cite{chaiken, smith72}.
This approach makes it possible to obtain smoothly 
previous 
polynomials 
as 
dehomogenizations of this universal
polynomial.
In other (closely related) language, these results provide essentially 
a projectively-weighted version of the matrix-tree theorem (cf.\ (\ref{pmtt})), extending the
results of \cite{maurer}. 

To present our results we need some additional background on the cut and cycle
matrices introduced in 
\ref{subsec-gen} (see \cite{bollobas} for further background). As before, we
focus on connected problems for simplicity.
It is well known 
that a square submatrix $A_T$
of $A$ of order $n-1$ 
has a non-null determinant if and only if the branches
defining the columns of $A_T$ form a spanning tree.
Moreover, for a given $A$ all such submatrices are known to have
the same determinant in absolute value;
based on this
we assign to $A$ a constant $k_A$, namely 
the 
positive integer
for which the identities $\det A_T = \pm k_A$ hold.
We have $k_A=1$ for so-called totally unimodular choices of $A$
(e.g.\ for a reduced incidence matrix). 
The same applies to maximal square submatrices of a fixed cycle matrix
$B$: non-singular ones
have a determinant of the form $\pm k_B$ for a positive integer
$k_B$, and this happens 
iff 
the chosen columns specify a co-tree (the complement
of a spanning tree), the determinant being zero otherwise.
Totally unimodular choices of $B$ are always possible.


With this background, the matrix-tree theorem in the unweighted context is simply 
expressed by the identity
$\det (AA^{\tra})=\tau k_A^2$, where $\tau$ is the total number of spanning
trees. If $A$ is totally unimodular
(in particular if it is a reduced incidence matrix), then the identity amounts
to $\det (AA^{\tra})=\tau$, which is probably the most popular form
of the matrix-tree theorem and which, in essence, can be traced back to
the 
work of
Kirchhoff and Maxwell \cite{kirchhoff1847, maxwell1892}. Dual results hold for $B$, namely, 
$\det (BB^{\tra})=\tau k_B^2$ holds true and, in particular, 
$\det (BB^{\tra})=\tau$ for totally unimodular choices of $B$. We will elaborate on 
these results (and their weighted counterparts) in subsection \ref{subsec-mtt}.



\subsection{Multihomogeneous Kirchhoff polynomial}


We assume in the sequel that the branch set $E(G)$ is simply 
$\mathbb{N}_m=\{1, \ldots, m\}$, so that the value assigned by the
source-free
configuration map $\hspace{0.5mm}\check{\hspace{-0.5mm}\gamma}$ to the $i$-th branch is 
$(p_i:q_i)$. Both $p_i$ and $q_i$ will be indeterminates
in the polynomial (\ref{kir-hom}); 
similarly,
$y_i=p_i/q_i$ and $z_i=q_i/p_i$ will arise as indeterminates
in certain dehomogenizations of this multihomogeneous polynomial.
Additionally, a spanning
tree will be represented by its set of branches; with
the convention above, 
a spanning tree
is unambiguously defined as a subset $T$ of $\mathbb{N}_m$
with $n-1$ elements. A cotree is denoted as $\overline{T}$,  
meaning $\mathbb{N}_m-T.$ By ${\cal T}$ we denote 
the set of spanning trees of the digraph. 


With this notation,
the multihomogeneous Kirchhoff polynomial of a connected
(di)graph (see \cite{chaiken, smith72} and references therein) is defined as
\begin{equation} \label{kir-hom}
K(p,q)=\sum_{T \in {\cal T}} 
\left( \prod_{j \in T} p_j \prod_{k \in \overline{T}} q_k\right),
\end{equation}
that is,
every
spanning tree $T$ in the graph sets up a monomial in $K(p,q)$,
which includes $p_i$ (resp.\ $q_i$) as a factor if the $i$-th branch
belongs to $T$ (resp.\ 
to 
$\overline{T}$). This polynomial 
is homogeneous (of degree one) in each pair of variables 
$(p_i, q_i)$, because necessarily either $p_i$ or $q_i$, but not
both, appears in each monomial; hence the ``multihomogeneous''
label.







\subsection{
Non-degenerate configurations}
\label{subsec-nondeg}


\begin{defin} \label{defin-nondeg}
A {\em non-degenerate configuration}
on a connected digraph $G$
is a configuration map 
$\gamma :E(G) \to \K\PP^2_*$ for which
the circuit equations (\ref{generalbranch}) have a unique
solution.
\end{defin}

\noindent {\bf Remark.} The non-degenerate character 
of a configuration does not depend on the excitation vector
$s$ (which only matters
for the location of the solution). We may hence study
the non-degeneracy of a configuration by examining the
associated source-free configuration
$\hspace{0.5mm}\check{\hspace{-0.5mm}\gamma}: E(G) \to \K\PP$
defined in subsection \ref{subsec-config} as 
$\pi \circ \gamma$.
This amounts in practice to the well-known fact that e.g.\ the voltage
of a 
voltage source does not matter for the existence and uniqueness of solutions, 
and its voltage can be hence fixed at zero. 
This is consistent with the fact that Theorem \ref{th-nondeg}
below can be stated in terms of source-free configurations. 
The other way round, the reader may understand 
that Theorem \ref{th-nondeg} applies to configurations with 
sources by assuming that
the polynomial $K$ depends vacuously on $s$: that is, if a 
pair of vectors $p$, $q$ annihilates the polynomial, then
so it does the triad $(p, q, s)$ for any $s$.





\begin{theor} \label{th-nondeg}
The set of (source-free) degenerate 
configurations  of a digraph
is the zero set of the multihomogeneous Kirchhoff polynomial (\ref{kir-hom}).
\end{theor}




The proof of Theorem \ref{th-nondeg} is based on several auxiliary
results that we state in advance. 
The first of them essentially says that 
the non-singularity of the coefficient matrix of the general model \eqref{generalbranch}
is equivalent to the one of the coefficient matrices of the reduced homogeneous models \eqref{homoggen}
and \eqref{homogsym}, namely
\begin{equation} \label{APBQ}
M=\begin{pmatrix} AP \\ BQ \end{pmatrix}.
\end{equation}

\begin{lema} \label{lema-corereduction}
Let $P$ and $Q$ be diagonal matrices in $\K^{m \times m}$
with $\begin{pmatrix}P & Q \end{pmatrix}$ of maximal rank.
If $A$ and $B$ are
arbitrary matrices in $\K^{r \times m}$ and $\K^{(m-r)\times m}$,
respectively, then
\begin{equation} \label{corereduction}
\det \begin{pmatrix} A & 0 \\ 0 & B \\ -Q & P \end{pmatrix}=
\det \begin{pmatrix} AP \\ BQ \end{pmatrix}.
\end{equation}
\end{lema}

\noindent {\bf Proof.} Write
\begin{equation}\label{identit1}
\begin{pmatrix} A & 0 \\ 0 & B \\ -Q & P \end{pmatrix}
\begin{pmatrix} P & -Q \\ Q & P \end{pmatrix}=
\begin{pmatrix} AP & -AQ \\ BQ & BP \\ 0 & P^2+Q^2 \end{pmatrix}
\end{equation}
because $P$ and $Q$ commute, and use the fact that 
\begin{equation} \label{identit2}
\det \begin{pmatrix} P & -Q \\ Q & P \end{pmatrix}=
\det (P^2+Q^2).
\end{equation}
The latter is very simple to check, since an obvious 
(determinant-preserving) permutation
of rows and columns
drives the matrix in the left-hand side to block-diagonal
form, with blocks of the form
\begin{equation*}
\begin{pmatrix} p_k & -q_k \\ q_k & \hspace{0.6mm}p_k \end{pmatrix}.
\end{equation*}
This makes 
it clear that the first determinant of (\ref{identit2})
amounts to 
\ifthenelse{\boolean{ieee}}{$\prod_{k=1}^m (p_k^2+q_k^2)=\det (P^2+Q^2)$.}{$$\displaystyle\prod_{k=1}^m (p_k^2+q_k^2)=\det (P^2+Q^2).$$}
Finally,
the maximal rank condition on $\PQ$ means 
that for each $k$ at least one of the two parameters $p_k$, $q_k$ 
does not vanish
and this makes both determinants in (\ref{identit2}) 
non-null. The result then follows
from the identity (\ref{identit1}).
\hfill $\Box$

\vv

Proposition \ref{propo-AB} below makes systematic use of certain 
results from linear algebra which are compiled here. Recall first that 
the {\em signature} of a permutation
is $(-1)^q$, where $q$ is the number of transpositions
in any decomposition of the permutation as a product of transpositions 
We are interested in certain classes of permutations of $m$ elements,
namely those in which 
$\mathbb{N}_m=\{1, \ldots, m\}$ can be partitioned in two
subsets $\sigma_1=\{j_1, \ \ldots \ j_r\}$,
$\overline{\sigma_1}=\mathbb{N}_m-\sigma_1=\{j_{r+1}, \ \ldots, j_m\}$
(where we assume $j_1 < j_2 < \ldots < j_r$ and
$j_{r+1} < \ldots < j_m$)
in a way such that the restrictions of 
the permutation to both $\sigma_1$ and $\overline{\sigma_1}$ 
are order-preserving. That is, if $k_i$ is the image of $j_i$ (for $i=1, \ldots, m$),
then both $k_1  < \ldots < k_r$ and $k_{r+1} < \ldots < k_m$ hold.
Letting $\sigma_2=\{k_1, \ \ldots \ k_r\}$ 
and $\overline{\sigma_2}=\{k_{r+1}, \ \ldots \ k_m\}$,
we will denote the permutation by 
${\cal P}_m(\sigma_1,\sigma_2)$,
allowed by the fact that it
is completely defined by 
$\sigma_1$ and $\sigma_2$; indeed, the elements of the
former are mapped into those of the latter in increasing order,
and the same happens with 
$\overline{\sigma_1}$ 
and $\overline{\sigma_2}$. 
In this context, the signature of the permutation can be computed as
follows (a detailed proof 
can be found e.g.\ in 
\cite{williamson2014}):
\begin{equation}\label{simplersign0}
\sgn({\cal P}_m(\sigma_1,\sigma_2))=(-1)^{\sum_{j \in \sigma_1}j+\sum_{k \in \sigma_2}k}.
\end{equation}

We will also use in Proposition \ref{propo-AB} the general
Schur complement (i.e.\ the 
Schur complement of a non-principal submatrix),
cf.\ \cite{generalSchur}.
Given $M \in \mathbb{K}^{m \times m}$
and two non-empty subsets $\alpha, \omega$
of $\mathbb{N}_m=\{1, \ldots, m\}$, 
denote by $M[\alpha, \omega]$
the submatrix of $M$ defined by the rows and columns specified
by the index sets $\alpha$ and $\omega$, respectively.
If $\alpha$ and $\omega$ have the same number
of elements and $M[\alpha, \omega]$ is non-singular, the
Schur complement of $\Ma$ in $M$ is defined as
$M/\Ma = M[\overline{\alpha},\overline{\omega}] - 
M[\overline{\alpha}, \omega](\Ma)^{-1}M[\alpha,\overline{\omega}],$
where again $\overline{\alpha}$, $\overline{\omega}$,
 stand for $\mathbb{N}_m-\alpha$, $\mathbb{N}_m-\omega$.
In this setting, we have the identity
\begin{equation} \label{schurdet}
\hspace{-2mm}\det M \hspace{-0.5mm}=\hspace{-0.5mm} \sgn({\cal P}_m(\alpha, \omega)) 
\det\hspace{-0.5mm}\left( \Ma \right) \det (M/\Ma).\hspace{-4mm}
\end{equation}



\begin{propo} \label{propo-AB}
Let $A$ and $B$ be two arbitrary cut and cycle matrices 
of a given connected digraph, with their 
columns arranged according to the same order of branches. Assume
that $T_1$
and $T_2$ 
specify 
two spanning trees,
and let $\overline{T}_1$ and $\overline{T}_2$ represent 
the corresponding cotrees.
Then
\begin{equation} \label{AB}
\det A_{T_1} \det B_{\overline{T}_1} = \sgn({\cal P}_m(T_1, T_2)) \det A_{T_2} \det B_{\overline{T}_2}.
\end{equation}
\end{propo}

\noindent {\bf Proof.} Let $\alpha=\{1, \ldots, n-1\}$. 
For notational brevity, denote by $A_{T_i}$ 
the submatrix of $A$ defined by the
columns indexed by $T_i$
(that is, $A_{T_i}=A[\alpha, T_i]$); analogously,
$B_{T_i}$ is the submatrix of $B$ defined
by the columns indexed by $T_i$.
By 
writing 
\begin{equation} \label{matrixAB}
\det \dsp\begin{pmatrix} A \\ B\end{pmatrix}
\end{equation}
in terms of $\det A_{T_1}$ and $\det A_{T_2}$ (using (\ref{schurdet})), 
we get
\begin{eqnarray*}
\ifthenelse{\boolean{ieee}}{
\sgn({\cal P}_m(\alpha, T_1))\det A_{T_1} \det(B_{\overline{T}_1}-B_{T_1}A_{T_1}^{-1} A_{\overline{T}_1}) = \hspace{2cm} \\
\sgn({\cal P}_m(\alpha, T_2))\det A_{T_2} \det(B_{\overline{T}_2}-B_{T_2}A_{T_2}^{-1} A_{\overline{T}_2}) \hspace{9mm}}
{\hspace{-3mm}
\sgn({\cal P}_m(\alpha, T_1))\det A_{T_1} \det(B_{\overline{T}_1}-B_{T_1}A_{T_1}^{-1} A_{\overline{T}_1}) = 
\sgn({\cal P}_m(\alpha, T_2))\det A_{T_2} \det(B_{\overline{T}_2}-B_{T_2}A_{T_2}^{-1} A_{\overline{T}_2})}
\end{eqnarray*}
and therefore
\begin{eqnarray} 
\ifthenelse{\boolean{ieee}}{\label{detlarg} \det A_{T_1} \det(B_{\overline{T}_1}-B_{T_1}A_{T_1}^{-1} A_{\overline{T}_1}) = \hspace{27mm} \nonumber \\
\sgn({\cal P}_m(T_1, T_2))\det A_{T_2} \det(B_{\overline{T}_2}-B_{T_2}A_{T_2}^{-1} A_{\overline{T}_2})\ }
{\det A_{T_1} \det(B_{\overline{T}_1}-B_{T_1}A_{T_1}^{-1} A_{\overline{T}_1}) = 
\sgn({\cal P}_m(T_1, T_2))\det A_{T_2} \det(B_{\overline{T}_2}-B_{T_2}A_{T_2}^{-1} A_{\overline{T}_2})\label{detlarg}}
\end{eqnarray}
since $\sgn({\cal P}_m(\alpha, T_1))\sgn({\cal P}_m(T_1, T_2))=\sgn({\cal P}_m(\alpha, T_2))$.

Making use of the fact that $\det B_{\overline{T}_i} \det B_{\overline{T}_i}^{\tra}=(\det B_{\overline{T}_i})^2=k_B^2$ 
does not depend on $i$, multiply the left-hand side of (\ref{detlarg}) 
by $\det B_{\overline{T}_1} \det B_{\overline{T}_1}^{\tra}$ 
and the right-hand side by $\det B_{\overline{T}_2}\det B_{\overline{T}_2}^{\tra}$ 
to recast this identity as
\ifthenelse{\boolean{ieee}}{\begin{eqnarray} 
&&\hspace{-13mm}\det A_{T_1} \det B_{\overline{T}_1} \det(B_{\overline{T}_1}B_{\overline{T}_1}^{\tra}-B_{T_1}A_{T_1}^{-1} A_{\overline{T}_1}B_{\overline{T}_1}^{\tra}) =
 \nonumber 
\\&&\hspace{-6mm}
\sgn({\cal P}_m(T_1, T_2))\det A_{T_2} \times \nonumber  
\\ &&
\times \det B_{\overline{T}_2} 
\det(B_{\overline{T}_2}B_{\overline{T}_2}^{\tra}-B_{T_2}A_{T_2}^{-1} A_{\overline{T}_2}B_{\overline{T}_2}^{\tra}). \ \ \label{detlarga0}
\end{eqnarray}}
{\begin{eqnarray} 
\det A_{T_1} \det B_{\overline{T}_1} \det(B_{\overline{T}_1}B_{\overline{T}_1}^{\tra}-B_{T_1}A_{T_1}^{-1} A_{\overline{T}_1}B_{\overline{T}_1}^{\tra}) =
\hspace{30mm} \nonumber \\
\sgn({\cal P}_m(T_1, T_2))\det A_{T_2} \det B_{\overline{T}_2} 
\det(B_{\overline{T}_2}B_{\overline{T}_2}^{\tra}-B_{T_2}A_{T_2}^{-1} A_{\overline{T}_2}B_{\overline{T}_2}^{\tra}). \label{detlarga0}
\end{eqnarray}
}

By writing the orthogonality 
of the cycle and cut spaces
as $AB^{\tra}=0$ 
one can derive the identity
$-A_{T_i}^{-1}A_{\overline{T}_i}B_{\overline{T}_i}^{\tra}=B_{T_i}^{\tra}.$ 
This allows us to
rewrite the left-hand side of (\ref{detlarga0}) as
\ifthenelse{\boolean{ieee}}{
\begin{eqnarray*} 
&& \hspace{-15mm} \det A_{T_1} \det B_{\overline{T}_1} \det(B_{\overline{T}_1}B_{\overline{T}_1}^{\tra}+B_{T_1}B_{T_1}^{\tra}) = \\
&& \hspace{20mm} \det A_{T_1} \det B_{\overline{T}_1} \det(BB^{\tra})
\end{eqnarray*}
and, analogously, the right-hand side as
\begin{eqnarray*} 
\sgn({\cal P}_m(T_1, T_2))\det A_{T_2} \det B_{\overline{T}_2} 
\det(BB^{\tra}). 
\end{eqnarray*}
}{
\begin{eqnarray*} 
\det A_{T_1} \det B_{\overline{T}_1} \det(B_{\overline{T}_1}B_{\overline{T}_1}^{\tra}+B_{T_1}B_{T_1}^{\tra}) =
\det A_{T_1} \det B_{\overline{T}_1} \det(BB^{\tra})
\end{eqnarray*}
and, analogously, the right-hand side as
\begin{eqnarray*} 
\sgn({\cal P}_m(T_1, T_2))\det A_{T_2} \det B_{\overline{T}_2} 
\det(BB^{\tra}). 
\end{eqnarray*}
}
The identity (\ref{AB}) then follows from the fact that 
$\det(BB^{\tra})=\tau k_B^2$ does not vanish. 
\hfill $\Box$

\vspace{3mm}

The key idea in our approach is that there is 
no chance to determine whether the signs of $\det A_{T_1}$ and
$\det A_{T_2}$ are the same in purely combinatorial terms, that is,
by just looking at
the permutation ${\cal P}_m(T_1, T_2)$. 
The same applies to $\det B_{\overline{T}_1}$ and $\det B_{\overline{T}_2}$.
But it is indeed possible to do it when we look at 
the {\em products} $\det A_{T_i}\det B_{\overline{T}_i}$, as shown above.
Another way to express the same is to say, allowed by
Proposition \ref{propo-AB}, that, for any pair of 
cut/cycle matrices $A$, $B$, the
quantity
\begin{equation} \label{kABconstant}
k_{AB}=(-1)^{\frac{n(n-1)}{2}+\sum_{j \in T} j}\det A_T \det B_{\overline T}
\end{equation}
is a (non-null)
invariant, that is, it does not depend on the actual choice of the spanning tree 
$T$. To prove this just use 
(\ref{simplersign0}), (\ref{AB}) and the fact that $(-1)^{\frac{n(n-1)}{2}}$ is a quantity
which does not depend on the tree $T$.

Accordingly, we will say that $A$ and $B$ are {\em positively matched} if 
$k_{AB}$ is positive
and, in particular, that they are {\em well-matched}
if $k_{AB}=1$.  Since $\det A_T$ and $\det B_{\overline T}$ are integers,
$A$ and $B$ are well-matched if and only if they are positively matched
and both of them are totally unimodular.
An important consequence (that the author could not find in the literature)
is that the fundamental matrices
\begin{equation} \label{fundamental}
A=\begin{pmatrix} I & A_{\overline{T}} \end{pmatrix}, \ 
B=\begin{pmatrix} B_{T} & I \end{pmatrix},\text{ with } B_T = -A_{\overline{T}}^{\tra} \text{,}
\end{equation}
defined by any spanning tree $T$ are well-matched since, on the one hand, 
$T=\{1, \ldots, n-1\}$, 
so that the exponent in (\ref{kABconstant}) is even, 
and on the other $\det A_T=\det B_{\overline T}=+1$ by construction
(both $A_T$ and  $B_{\overline T}$ are identity matrices). 
We will elaborate on the implications of this later (cf.\ Corollary \ref{coro-mtt2}).

\begin{versionC}
\color{red}For the moment let us mention that, in order to check if two arbitrary matrices $A$, $B$ are
positively matched, it is enough to fix a tree/cotree pair and
check if the sign of $\det A_T \det B_{\overline T}$ is consistent
with the parity of $n(n-1)/2+\sum_{j \in T} j$. Things get simpler
if the tree branches are chosen as the first $n-1$ ones, since
in this case the later sum is even and the problem amounts
to checking whether $\det A_T$ and $\det B_{\overline T}$ have the same sign.
\color{black}
\end{versionC}

\vspace{4mm}

\noindent {\bf Proof of Theorem \ref{th-nondeg}.} 
The matrix in the left-hand side of (\ref{corereduction})
is the coefficient matrix of 
\eqref{generalbranch}
and therefore the 
characterization of non-degenerate configurations
amounts to characterizing the non-singularity of this matrix.
To do this, we compute this determinant 
using a generalized Laplace expansion 
along the first $n-1$ rows of the matrix (\ref{APBQ}), arising
in the right-hand side of (\ref{corereduction}).
Generalized Laplace expansions 
are explained  in many linear algebra texts; see e.g.\ \cite{loehr}.
With $\alpha=\{1, \ldots, n-1\}$, 
such expansion 
of $\det M$ 
reads as
\begin{equation*}
(-1)^{n(n-1)/2} \sum_{|\omega|=n-1} (-1)^{\sum_{j \in \omega}j}
\det (\Ma) \det (\Mo),
\end{equation*}
the exponent $n(n-1)/2$ being 
the sum of the indices
of $\alpha$.

Because of the diagonal structure of $P$ and $Q$ and 
the properties of the digraph matrices $A$ and 
$B$ stated at the beginning of this section, the non-null
determinants of the submatrices arising in this expansion
come from spanning tree/cotree pairs
and have the form
$\det A_T \det P_T$ (where $P_T$ is the submatrix of $P$ 
defined by the rows and columns indexed by $T$) 
and $\det B_{\overline{T}} \det Q_{\overline{T}}$
(with the same 
convention for $Q_{\overline{T}}$; find
the notation for $A_T$, $B_{\overline{T}}$ in the proof of Proposition
\ref{propo-AB}).
Using these remarks,
we may recast the expansion above as
\begin{equation*}
(-1)^{n(n-1)/2} \sum_{T \in {\cal T}} (-1)^{\sum_{j \in T}j}
\det A_T \det P_T 
\det B_{\overline{T}} \det Q_{\overline{T}}.
\end{equation*}
Now, the key step in the proof is the fact that 
\begin{equation}\label{invariant}
(-1)^{n(n-1)/2} (-1)^{\sum_{j \in T}j}
\det A_T 
\det B_{\overline{T}}
\end{equation}
does not depend on the choice of the tree $T$
(this is the constant $k_{AB}$ in \eqref{kABconstant}). 
Hence
\ifthenelse{\boolean{ieee}}{
\begin{eqnarray*} 
\hspace{-1mm}\det 
\begin{pmatrix} AP \\ BQ \end{pmatrix}=
k_{AB} \hspace{-1mm}\sum_{T \in {\cal T}}\hspace{-0.5mm} \det P_T \det Q_{\overline{T}} 
=k_{AB}\hspace{-1mm}\sum_{T \in {\cal T}} \hspace{-0.5mm}
\left( \prod_{i \in T} p_i \prod_{j \in \overline{T}} q_j\right) 
\end{eqnarray*}
}{
\begin{equation*} 
\det 
\begin{pmatrix} AP \\ BQ \end{pmatrix}
= k_{AB} \sum_{T \in {\cal T}} \det P_T \det Q_{\overline{T}} =
k_{AB}\sum_{T \in {\cal T}} 
\left( \prod_{i \in T} p_i \prod_{j \in \overline{T}} q_j\right), 
\end{equation*}
}
that is,
\begin{equation} \label{equimp}
\det 
\begin{pmatrix} AP \\ BQ \end{pmatrix}
=k_{AB} K(p,q),
\end{equation}
and the claim follows from the fact that $k_{AB} \neq 0$.
\hfill $\Box$




\vv

The following is essentially the unweighted version of Theorem 
\ref{th-nondeg}, which is obtained by setting
$P=Q=I$ in (\ref{equimp}); 
just note that
$K(\mathbf{1},\mathbf{1})=\tau$, where $\mathbf{1}=(1, \ldots, 1)$.

\begin{coro}\label{coro-mtt1}
For any pair of cut/cycle matrices $A$, $B$, we have
\begin{equation}\label{mmt0}
\det \begin{pmatrix}A \\ B\end{pmatrix} = \tau \hspace{0.5mm}k_{AB},
\end{equation}
where $\tau$ is the number of spanning trees and $k_{AB}$ is the constant arising in (\ref{kABconstant}).
\end{coro}

At this point it is worth comparing (\ref{mmt0})
with previous (and closely related) results in this direction, 
a full account of which can be found in \cite{chen,maurer}.
Chen's approach obtains for the determinant \eqref{mmt0} 
the expression 
$\pm \tau k_A k_B$: this is, specifically, (2.144b) in \cite{chen}
(essentially the same holds in \cite{maurer}, cf.\ Theorem 4' there),
but an ambiguity in the sign remains. 
It is clear 
that $k_{AB}=\pm k_A k_B$ 
but we emphasize 
that with our approach $k_{AB}$ captures the $\pm$ sign,
which can be computed from any tree-cotree pair using (\ref{kABconstant}).
Actually, the difference between
both approaches is a subtle one: Chen uses an 
indirect approach to compute {\em up to a sign} the determinant
(\ref{mmt0}), and then derives a result
(Lemma 2.8 in \cite{chen}) which can be understood as an alternative
statement of
Proposition \ref{propo-AB} above. By contrast, our approach 
captures the essential combinatorial property expressed by (\ref{AB})
in Proposition \ref{propo-AB},
to derive from it the expression 
\eqref{mmt0} 
with a well-defined sign.

\begin{coro} \label{coro-mtt2}
If $A$ and $B$ 
are well-matched cut/cycle matrices
(in particular, if they are the fundamental matrices (\ref{fundamental})
defined by a spanning tree), then 
\begin{equation}\label{mtt}
\det \begin{pmatrix} A \\ B \end{pmatrix} = \tau.
\end{equation}
\end{coro}


\noindent This result,
which provides
an alternative form
of the matrix-tree theorem giving
the number of spanning trees 
in terms of (well-matched) cut/cycle matrices, 
is just (\ref{mmt0}) with $k_{AB}=1$. 


\subsection{Dehomogenization. Classical forms of the
Kirchhoff polynomial and the matrix-tree theorem}
\label{subsec-mtt}

For simplicity, assume in what follows $k_{AB}=1$, namely, that
$A$ and $B$ are well-matched. Then (\ref{equimp}) reads as
\begin{equation}\label{pmtt}
\det \begin{pmatrix}AP \\ BQ\end{pmatrix} = K(p,q).
\end{equation}

From this expression, which can be understood as a projectively-weighted
version of the matrix-tree theorem,
we may derive certain known forms
of this theorem in a weighted setting,
involving classical forms of the Kirchhoff polynomial which are shown below
to arise as dehomogenizations of (\ref{kir-hom}). 


Let us focus the attention on the $\mathbb{A}_{\check{\mathbf{y}}}$ affine patch
defined in $\K\PP \times \ldots \times \K\PP$ by the 
conditions
$q_i \neq 0$ for $i=1, \ldots, m$. This is the patch where the 
admittance matrix $Y$ is well-defined (and equals $PQ^{-1}$ for
arbitrary choices of $P$, $Q$, as far of course as $Q$ is non-singular).
In this patch the choice $\hat{Q}=I$ is always possible;
this yields $\hat{P}=Y$.
With these parameters,
the 
matrix in (\ref{pmtt})
amounts to the one in the branch-voltage
system (\ref{branchvoltage}), and the identity (\ref{pmtt}) becomes
\begin{equation} \label{maurer1}
\det 
\begin{pmatrix} AY \\ B \end{pmatrix}
=K(y,\mathbf{1}),
\end{equation}
(cf. \cite{maurer} in this regard). 
The polynomial in the
right-hand side  of (\ref{maurer1})
is a dehomogenization of $K(p,q)$,
which amounts to the so-called
Maxwell's form or tree-based form of the 
Kirchhoff polynomial, namely,
\begin{equation} \label{kir-trees}
K_0(y)=\sum_{T \in {\cal T}} \prod_{j \in T} y_j,
\end{equation}
which is set up simply by inserting
the admittance parameter $y_j$ in the monomial corresponding to
the spanning tree $T$ if
branch $j$ belongs to $T$.


The vectors of admittances which do
not annihilate the polynomial (\ref{kir-trees}) define the set
of non-degenerate
configurations in the voltage-controlled patch
$\mathbb{A}_{\check{\mathbf{y}}}$; equivalently,
these are the admittances for which the branch-voltage system
(\ref{branchvoltage}) (which is defined only on 
$\mathbb{A}_{\check{\mathbf{y}}}$) has a unique solution. 
Moreover, from (\ref{maurer1})
it is not difficult to derive also the identity
\begin{equation} \label{nodal}
\det \left(AY\hspace{-1mm}A^{\tra}\right) = K_0(y),
\end{equation}
which is the usual form of the weighted matrix-tree theorem
(note, in particular, that $A$ may be a reduced
incidence matrix). 
This shows that the non-vanishing of $K_0(y)$ also characterizes
the set of admittances where the classical nodal equations are well-defined.
Note finally that by setting $Y=I$ in (\ref{nodal}) we get the unweighted version
of the matrix-tree theorem in its classical form, 
namely $\det \left(A
A^{\tra}\right) = \tau$: here
$A$ typically denotes
a reduced incidence matrix.

We leave it to the reader to check that the 
expression for the 
polynomial which 
characterizes non-degenerate configurations in the patch
$\mathbb{A}_{\check{\mathbf{z}}} \subseteq \K\PP \times \ldots \times \K\PP$ is
\begin{equation} \label{kir-cotrees}
K_1(z)=\sum_{T \in {\cal T}} \prod_{k \in \overline{T}} z_k,
\end{equation}
arising as the dehomogenization $K(\mathbf{1}, z)$. 
The zeros of this polynomial characterize the 
degeneracies of the branch-current system (\ref{branchcurrent}) and also 
of the 
loop analysis equations. In this case the polynomial is constructed
by including, in the monomial corresponding
to a given tree, the impedance parameter $z_k$ iff the $k$-th branch
belongs to the cotree. 
And even if we omit it for brevity, 
it is possible to derive analogously 
the polynomial characterizing
non-degenerate configurations in so-called hybrid models, 
mixing admittance and impedance descriptions 
(find a detailed
discussion in this regard in \cite{GR14}). 

The form of the different polynomials arising above can be easily illustrated
in terms of the 
graph in Fig.\ 2(a). 
It is obvious that there are three spanning trees, 
each one excluding one of the branches.
The universal Kirchhoff polynomial \eqref{kir-hom} 
has in this case the expression
depicted in \eqref{polyn1}: here the term $p_1p_2q_3$ comes from the tree defined
by branches 1 and 2, which are responsible for the $p_1p_2$ factor, whereas
branch 3 defines the co-tree and yields the $q_3$ factor. The other
terms are obtained analogously.
In turn, the form \eqref{kir-trees} reads here $y_1y_2+y_1y_3+y_2y_3$, 
each term
coming from one of the spanning trees, and is only valid if all
branches admit an admittance description
(namely, this holds in the patch $\mathbb{A}_{\check{\mathbf{y}}}$). 
The dual case is given by
\eqref{kir-cotrees} and reads $z_1+z_2+z_3$ (cf.\ Section \ref{sec-intro}), 
where each term arises in
this case as the impedance in each co-tree; this expression is
valid only in the patch $\mathbb{A}_{\check{\mathbf{z}}}$.




\section{Controlled sources}
\label{sec-controlled}

The framework developed in previous sections can be extended to accommodate 
controlled sources, as discussed below. 
A salient advantage of our approach in this context is that
it avoids the need for the classical distinction among the four types of sources
(depending on the controlling/controlled variables); independent sources
and switches 
can be also included in a comprehensive manner. For the sake of brevity we only 
sketch the results.

Disregarding excitation terms, an abstract controlled source is described by a pair of equations 
of the form
\begin{subequations} \label{CSs}
\begin{eqnarray}
p_1 v_1 - q_1 i_1 & = & 0 \\
p_2 v_2 - q_2 i_2 + \alpha v_1 + \beta i_1 & = & 0,
\end{eqnarray}
\end{subequations}
with $(p_k, q_k) \neq (0, 0)$ for $k=1, 2$. The first (resp.\ second) equation describes the controlling
(resp.\ controlled) device. Ideal VCVS's (voltage-controlled voltage sources), CCVS's, VCCS's
and CCCS's (with the same convention in the acronyms) 
are described by the parameter values $q_2=0$, $\alpha \neq 0$, $\beta =0$;
$q_2=0$, $\alpha=0$, $\beta \neq 0$; $p_2=0$, $\alpha \neq 0$, $\beta =0$; and
$p_2=0$, $\alpha=0$, $\beta \neq 0$, respectively.
Non-ideal cases, displaying an impedance in series or parallel 
with an ideal voltage/current source,
are included above with 
$p_2 \neq 0 \neq q_2$, with the Th\'evenin/Norton equivalence holding again. The addition
of excitation terms $s_1$ and $s_2$ in the right-hand side would easily accommodate independent
sources in the same setting. Note that the formalism above 
may also account 
for ideal switches, since the parameter values 
$q_2=0$ and $p_2=0$, respectively, model
a closed switch (short-circuit) and an open switch, 
with $\alpha=\beta=0$ in both cases;
this might of interest, for instance, in modelling saturation/cut-off regimes in transistors.

\begin{figure}[ht] \label{fig-smallsignal}
\vspace{1.5mm}
\ifthenelse{\boolean{ieee}}{\hspace{13mm}}{\hspace{15mm}} 
\ifthenelse{\boolean{ieee}}{\epsfig{figure=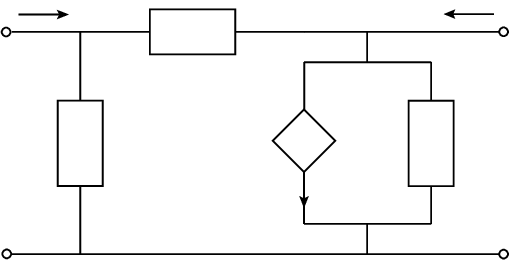, width=0.33\textwidth}}
{\hspace{33mm}\epsfig{figure=smallsignal_bis.eps, width=0.4\textwidth}}
\ifthenelse{\boolean{ieee}}{
\put(-160,86){\small $i_{\mathrm{in}}$}
\put(-182,37){\small $v_{\hspace{0.2mm}\mathrm{in}}$}
\put(-180,74){\footnotesize $+$}
\put(-180,0){\footnotesize $-$}
\put(-18,86){\small $i_{\mathrm{out}}$}
\put(3,74){\footnotesize $+$}
\put(3,0){\footnotesize $-$}
\put(-1,37){\small $v_{\hspace{0.2mm}\mathrm{out}}$}
\put(-144.7,37){\small $1$}
\put(-107,73.5){\small $3$}
\put(-72.5,37){\small $2a$}
\put(-30.5,36.5){\small $2b$}
}{
\put(-179,96){\small $i_{\mathrm{in}}$}
\put(-201,41){\small $v_{\hspace{0.2mm}\mathrm{in}}$}
\put(-199,82){\footnotesize $+$}
\put(-199,0){\footnotesize $-$}
\put(-22,96){\small $i_{\mathrm{out}}$}
\put(3,82){\footnotesize $+$}
\put(3,0){\footnotesize $-$}
\put(-1,41){\small $v_{\hspace{0.2mm}\mathrm{out}}$}
\put(-160.5,40){\small $1$}
\put(-119,81){\small $3$}
\put(-81,40.7){\small $2a$}
\put(-34,40){\small $2b$}
}
\caption{Small-signal $\Pi$-model of a transistor.}
\end{figure}
\vspace{-1mm}

The interest of this formulation relies on the fact that a unique analysis can be performed
for all four types of controlled sources 
and accounting also for all possible open- and short-circuits 
in the same model. 
We illustrate this idea by considering an
abstract $\Pi$-model of the small-signal equivalent of a transistor depicted in
Fig.\ 3. Here, the abstract controlled source (which is controlled by the first branch)
would amount to a CCCS for
a bipolar junction transistor and to a VCCS for a 
MOSFET, but we do not need to 
make such a distinction. 
We join together the 
parallel impedance (2b in the figure) with the controlled source 
in a single
circuit element. 
To be specific, we focus on the problem of the
existence of a two-port, $Z$-parameter description. 

Using $v_1=v_{\mathrm{in}}$ and $v_2=v_{\mathrm{out}}$ 
the circuit equations can be written as 
\begin{subequations}
\begin{eqnarray}
i_{\mathrm{in}} & = & i_1 + i_3 \\
i_{\mathrm{out}} & = & i_2 - i_3 \\
v_1 & = & v_2 + v_3 \\
p_1 v_1 - q_1 i_1 & = & 0 \\
p_2 v_2 - q_2 i_2 + \alpha v_1 + \beta i_1 & = & 0\\
p_3 v_3 - q_3 i_3 & = & 0,
\end{eqnarray}
\end{subequations}
and the existence of a $Z$-parameter description relies on the non-singularity of the coefficient
matrix of the variables $v_1=v_{\mathrm{in}}$, $v_2=v_{\mathrm{out}}$, $v_3$,
$i_1$, $i_2$ and $i_3$, since this makes it possible to write $v_{\mathrm{in}}$ and $v_{\mathrm{out}}$ 
just in terms of $i_{\mathrm{in}}$ and $i_{\mathrm{out}}$, as intended. An easy computation
shows that the determinant of this 
coefficient matrix is 
\begin{equation} \label{detCS}
p_1 p_2 q_3 + p_1 q_2 p_3 + q_1 p_2 p_3 + \alpha q_1 p_3 +\beta  p_1 p_3,
\end{equation}
the non-vanishing of which characterizes the existence of a $Z$-parameter description.
From the general expression (\ref{detCS}) one may draw conclusions in many different settings. 
Just to give a glimpse, let us assume that the controlled source is an ideal 
current source, so that $p_2=0$, and that the bridge admittance 
does not vanish (that is, $p_3 \neq 0$), 
in order to examine the dependence of the non-degeneracy expression above on the controlling
branch. Under these assumptions, the vanishing of (\ref{detCS}) amounts to that of 
$p_1q_2 +  \alpha q_1 + \beta p_1.$
In the CCCS context ($\alpha = 0$; the BJT case) this further amounts to $p_1 q_2 + \beta p_1 $,
whereas in the VCCS one ($\beta =0$; the MOSFET case) the expression reads as $p_1 q_2 +  \alpha q_1$.
Should the controlling branch be a (say) regular impedance ($p_1 \neq 0 \neq q_1$) then
both contexts are essentially the same, since one can easily resort from the voltage-controlled 
to the current-controlled framework and vice-versa, just setting the gains in a way such 
that $\alpha q_1=\beta p_1$. More can be derived from the homogeneous formalism, though: indeed,
when the controlling branch is open-circuited ($p_1=0$) then the BJT case always degenerates and no
$Z$-description holds; by contrast, the assumption $p_1=0$ poses no problem for the MOSFET, 
since there is a non-null extra term of the form $\alpha q_1 $. We emphasize the fact
that these conclusions can be derived from a single model; needless to say, they can be also 
obtained
from classical circuit analysis but one would need to set up 
different models in order to cover the variety of scenarios.

Several remarks are in order regarding the extension of the results of Sections \ref{sec-homog} and \ref{sec-nondeg},
even if a detailed analysis is left for future work. Note that
we may account for coupling parameters in the $P$ and $Q$ matrices simply by 
writing the blocks corresponding
to (\ref{CSs}) as
$$P = \begin{pmatrix} p_1 & 0 \\ \alpha & p_2 \end{pmatrix}, \
Q = \begin{pmatrix} q_1 & 0 \\ -\beta & q_2 \end{pmatrix}.$$
One can easily check that the matrix $\begin{pmatrix}P & Q\end{pmatrix}$ (or, equivalently, $\begin{pmatrix}-Q & P\end{pmatrix}$) 
still has maximal rank 
and this means that for each controlling-controlled pair
the characteristic equation
$-Qi+Pv=0$ (that is, (\ref{CSs})) can be again described for all possible 
parameter values just in terms of two homogeneous
variables $u_1$, $u_2$: such a parametrization
now reads as 
$i = \hat{P}u$, $v= \hat{Q}u$,
with $u=(u_1, u_2)$ and blocks of the form
\begin{equation} \label{auxcontrolled}
\hat{P}= \begin{pmatrix} p_1 & 0 \\ \gamma & p_2 \end{pmatrix}, \ 
\hat{Q} = \begin{pmatrix}
q_1 & 0 \\ \delta & q_2 \end{pmatrix}, 
\end{equation} 
where
$\displaystyle\gamma = \frac{q_2}{p_2^2+q_2^2}
\left(\alpha q_1 + \beta p_1\right), \ \delta = \frac{-p_2}{p_2^2+q_2^2}
\left(\alpha q_1 + \beta p_1\right). 
\vspace{3mm} $
This is possibly of interest in order to 
include in different circuit models
all possible types of controlled sources
in terms of a single (and the same) pair of variables $u_1$, $u_2$. 
Moreover, and even if we state the following without proof, 
an identity such as (\ref{corereduction}) within the reduction
process arising in Lemma \ref{lema-corereduction} 
still holds in this context
when $P$ and $Q$ in the right-hand side of (\ref{corereduction}) 
are replaced by $\hat{P}$ and $\hat{Q}$ as defined above.
We also note that the terms involving control parameters (i.e.\ $\alpha$, $\beta$)
in determinantal expansions such as 
(\ref{detCS}) may be addressed in terms of certain pairs of spanning trees,
accounting for off-diagonal terms in the $P$ and $Q$ matrices. 
Finally, fully coupled problems are also in the scope of future research.
More difficulties are likely to show up in this context, including not only analytical but also computational
aspects; worth mentioning in this regard is the term cancellation problem: cf.\ \cite{symbolic}.


\section{Example: fault isolation in a bridge circuit}
\label{sec-faults}


We illustrate in this section, by means of a simple example, how 
the above approach can be used in practice. 
To this end, consider the Wien bridge
circuit depicted in Fig.\ 4(a). 

\ifthenelse{\boolean{ieee}}{
\begin{figure}[ht] \label{fig-faults}
\hspace{-7mm}
\parbox{0.5in}{
\epsfig{figure=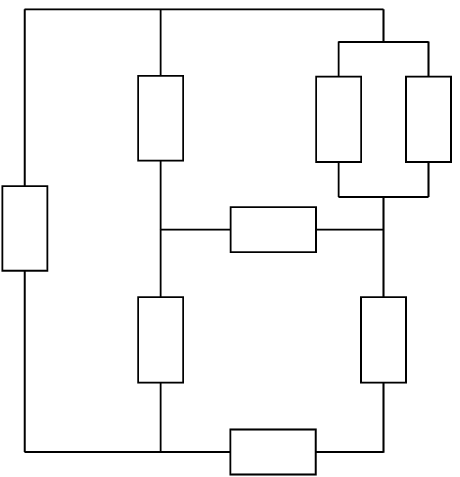, width=0.235\textwidth}
\put(-116.2,62.5){{\small 0}} 
\put(-80,92.6){{\small 1}}
\put(-80,33){{\small 2}}
\put(-35,92.3){{\small 3a}}
\put(-11.5,92.3){{\small 3b}}
\put(-23,33){{\small 4a}}
\put(-53.5,4){{\small 4b}}
\put(-50.2,62.4){{\small 5}}
}
\vspace{2mm}
\hspace{30mm}
\parbox{0.5in}{
\epsfig{figure=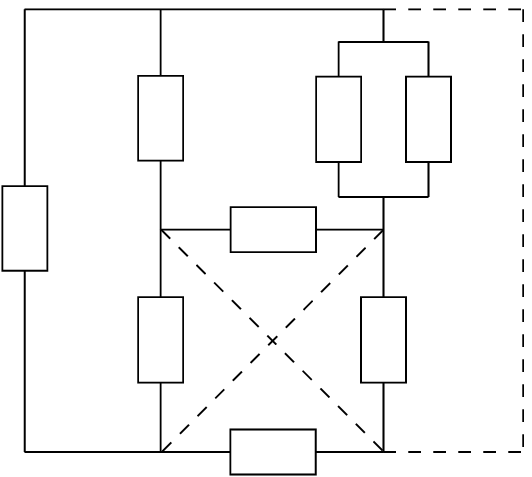, width=0.27\textwidth}
\put(-134,61.5){{\small 0}}
\put(-98,91.8){{\small 1}}
\put(-98.3,33){{\small 2}}
\put(-54,91){{\small 3a}}
\put(-30.3,91){{\small 3b}}
\put(-42,32.4){{\small 4a}}
\put(-72,4){{\small 4b}}
\put(-69.3,62){{\small 5}}
}
\caption{(a) Wien bridge \ \ \ \ \ \ 
(b) Virtual branches modelling bridging  faults.} 
\end{figure}
}
{
\begin{figure}[ht] \label{fig-faults}
\vspace{5mm}
\hspace{0mm}
\parbox{0.5in}{
\epsfig{figure=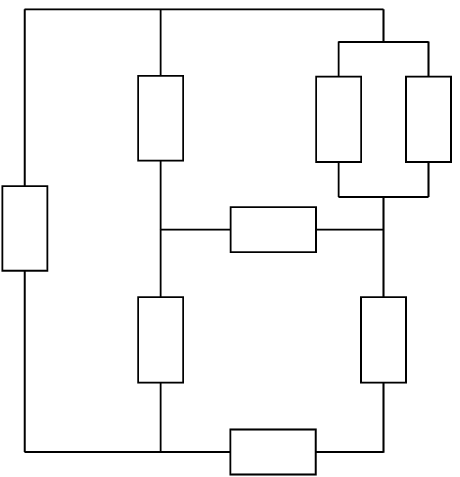, width=0.38\textwidth}
\put(-171.5,93){{\small 0}}
\put(-119,136){{\small 1}}
\put(-119,50){{\small 2}}
\put(-51,136){{\small 3a}}
\put(-16,136){{\small 3b}}
\put(-33,50){{\small 4a}}
\put(-76.5,6.5){{\small 4b}}
\put(-74,94){{\small 5}}
}
\vspace{2mm}
\hspace{70mm}
\parbox{0.5in}{
\epsfig{figure=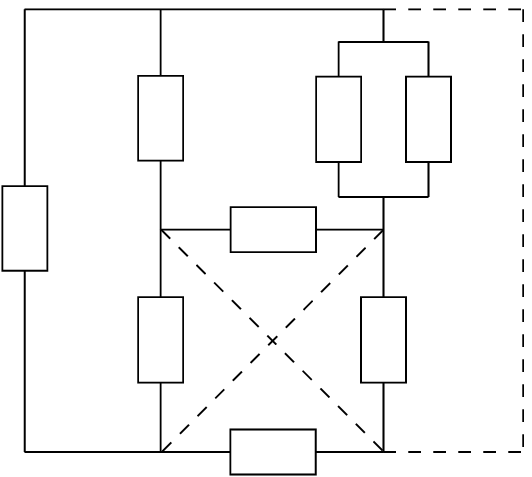, width=0.44\textwidth}
\put(-200,93){{\small 0}}
\put(-146.5,136){{\small 1}}
\put(-146.5,50){{\small 2}}
\put(-80,136){{\small 3a}}
\put(-45,136){{\small 3b}}
\put(-62,50){{\small 4a}}
\put(-106,6.5){{\small 4b}}
\put(-103,94){{\small 5}}
}
\caption{(a) Wien bridge \ \ \ 
\ \ \ \ \ \ \ \ \ \ \
\ \ \ \ \ \ \ \ \ \ \ \ 
(b) Virtual branches modelling bridging  faults.} 
\end{figure}
}

\mbox{}\vspace{-5mm}

In a classical approach, we may assume that
all linear elements are impedances 
(exception made of the 0-th branch, 
which will correspond to an ideal voltage source with voltage
$v_0$) and that they are
characterized by their impedance parameter $z_k$. To fix simple
values, let the bridge be balanced 
(namely, let $z_1/z_2=z_3/z_4$,
with $z_3$ and $z_4$ denoting the parallel/series connection
of $z_{3a}$ and $z_{3b}$ and of $z_{4a}$ and $z_{4b}$) 
by taking e.g.\ $z_1=1,$
$z_2=2$,
$z_{3a}=z_{4a}=1,$ $z_{3b}=z_{4b}=-j$; 
additionally, in order to model a substantially larger impedance across
the bridge set e.g.\ $z_5=100$. 
Assume if you want that all impedances are in 
k$\Omega$'s. 
The balance condition 
yields $v_5=0$
regardless of the 
values of $z_5$ and $v_0$. 

Assume now that 
we want to simulate bridging faults in this circuit,
by checking the actual values displayed by $v_5$ 
in different (and possibly faulty) scenarios.
Bridging faults
arise e.g.\ in integrated circuits 
from spurious connections between metal interconnects, and are typically modelled as
short-circuits. In practice (in larger scale circuits) one may identify critical 
pairs of nodes which are sensitive to this type of faults and then 
simulate, for later use, the behavior of the faulty circuits resulting from
such bridging effects. According to this strategy, the goal in our context
would be to set up a table 
of expected values for $v_5$ under different faulty conditions,
making it possible to identify later
the spurious connection which would eventually be responsible for a
failure.

To perform such simulations, in a classical framework (say e.g.\
in terms of the branch-current model (\ref{branchcurrent})) one would need
to reformulate the topology of the circuit and hence
the model for each fault scenario, by adding a 
(short-circuit) virtual branch, as in Fig.\ 4(b). Note that
an additional current variable $i_k$ must be included in the
equations 
when modelling a fault in the $k$-th 
branch.
Be aware of the fact that the no-faults scenario cannot
be accommodated as a particular case of any of the resulting faulty models,
since both cases correspond to the extremal values $y_k=0$ (no fault)
and $z_k=0$ (bridging fault) and each model involves a different set of 
variables: specifically, the $i_k$ variable in the faulty scenario
for branch $k$ would not be present in the original model.

By contrast, if we resort to the homogeneous framework we may include
all faulty circuits and the no-fault original one in single model, even
accommodating simultaneous faults.
To this end, we may simply characterize the virtual branches  of Fig.\ 4(b)
by using homogeneous impedances $(p_k:q_k)$ for $k=$ 6, 7, 8
and homogeneous variables $u_6$, $u_7$ and $u_8$
(branches 6 and 7 are the NW-SE and SW-NE diagonals in the figure,
and branch 8 is the one on the right). Currents and voltages
will be computed as $i_k=p_ku_k$, $v_k=q_ku_k$ for $k=$ 6, 7, 8.
The resulting model 
\ifthenelse{\boolean{ieee}}{is depicted in (\ref{eq-faults1}).}{reads as}
\begin{figure*}[htb] 
\begin{equation}
\label{eq-faults1}
\left(\begin{array}{ccccccccccc} 
1 & 1 & 0 & 1 & 1 & 0 & 0 & 0 & 0 & 0 & p_8\\ 
0 & -1 & 1 & 0 & 0 & 0 & 0 & 1 & p_6 & 0 & 0\\ 
-1 & 0 & -1 & 0 & 0 & 0 & 1 & 0 & 0 & p_7 & 0\\ 
0 & 0 & 0 & -1 & -1 & 1 & 0 & -1 & 0 & -p_7 & 0\\ 
0 & z_1 & z_2 & 0 & 0 & 0 & 0 & 0 & 0 & 0 & 0\\ 
0 & 0 & 0 & -z_{3a} & z_{3b} & 0 & 0 & 0 & 0 & 0 & 0 \\ 
0 & 0 & 0 & -z_{3a} & 0 & -z_{4a} & z_{4b} & 0 & 0 & 0 & 0\\ 
0 & z_1 & 0 & -z_{3a} & 0 & 0 & 0 & z_5 & 0 & 0 & 0\\ 
0 & z_1 & 0 & -z_{3a} & 0 & -z_{4a} & 0 & 0 & q_6 & 0 & 0\\ 
0 & 0 & 0 & -z_{3a} & 0 & 0 & 0 & 0 & 0 & q_7 & 0\\ 
0 & 0 & 0 & -z_{3a} & 0 & -z_{4a} & 0 & 0 & 0 & 0 & q_8
\end{array} \right)
\begin{pmatrix}
i_0 \\
i_1 \\
i_2 \\
i_{3a} \\
i_{3b} \\
i_{4a} \\
i_{4b} \\
i_5 \\
u_6\\
u_7 \\
u_8
\end{pmatrix} = \begin{pmatrix}
0 \\
0 \\
0 \\
0 \\
v_0 \\
0 \\
-v_0 \\
0 \\
0 \\
-v_0\\
0 
\end{pmatrix}.
\end{equation}
\ifthenelse{\boolean{ieee}}{\vspace{0.5mm}\\\rule{\textwidth}{0.1 mm}\\\vspace{-6mm}\\}{}
\end{figure*}

To model the no-fault case just set all $p_k$'s to zero
(and, if you want, $q_k=1$ for simplicity, although this is not
strictly necessary), 
whereas the different bridging faults are obtained
just by resetting the corresponding parameter 
$q_k$ to $0$ (and, optionally, $p_k=1$). Here we need
to set up one model, instead of four, one for each individual fault
plus the original one, 
in the branch-current setting.
Worth emphasizing is the 
partially
homogeneous 
nature of (\ref{eq-faults1}),
where some branches are modelled in classical terms as
impedances (or as a source), 
with their 
current variables entering the model,
and only certain branches
are given 
a homogeneous description.

The same idea may be further exploited, e.g.\ to include
faults also in the original impedances (namely, $1$ to $4b$;
we do not model faults in the source or in $z_5$). 
Short-circuit faults at 
these branches can be naturally framed in the 
branch-current
model by setting $z_k=0$, but this is not the case
with open-circuit faults, resulting
e.g.\ from wire breaks, an excess of insulating material, etc. They 
should be modelled by the conditions
$y_k=0$ and again this would require 
another model; be also
aware that resorting to an admittance description in the original
model would exclude short-circuit faults.

Instead, by using 
homogeneous variables for all 
branches we accommodate
all scenarios in a single model, just setting $q_k$ or $p_k$ to zero
to model short-circuit and open-circuit faults. 
In this case we avoid defining up to
ten models in a classical framework 
by
setting up only
one model in the homogeneous context, namely
\ifthenelse{\boolean{ieee}}{the one in (\ref{eq-faults2}).}{}
\begin{figure*}[htb] 
\begin{equation}\label{eq-faults2}
\hspace{-6mm}\left(\begin{array}{ccccccccccc} 
p_0 & p_1 & 0 & p_{3a}  & p_{3b} & 0 & 0 & 0 & 0 & 0 & p_8\\ 
0 & -p_1 & p_2 & 0 & 0 & 0 & 0 & p_5 & p_6 & 0 & 0\\ 
-p_0 & 0 & -p_2 & 0 & 0 & 0 & p_{4b} & 0 & 0 & p_7 & 0\\ 
0 & 0 & 0 & -p_{3a} & -p_{3b} & p_{4a} & 0 & -p_5 & 0 & -p_7 & 0\\ 
-q_0 & q_1 & q_2 & 0 & 0 & 0 & 0 & 0 & 0 & 0 & 0\\ 
0 & 0 & 0 & -q_{3a} & q_{3b} & 0 & 0 & 0 & 0 & 0 & 0 \\ 
q_0 & 0 & 0 & -q_{3a} & 0 & -q_{4a} & q_{4b} & 0 & 0 & 0 & 0\\ 
0 & q_1 & 0 & -q_{3a} & 0 & 0 & 0 & q_5 & 0 & 0 & 0\\ 
0 & q_1 & 0 & -q_{3a} & 0 & -q_{4a} & 0 & 0 & q_6 & 0 & 0\\ 
q_0 & 0 & 0 & -q_{3a} & 0 & 0 & 0 & 0 & 0 & q_7 & 0\\ 
0 & 0 & 0 & -q_{3a} & 0 & -q_{4a} & 0 & 0 & 0 & 0 & q_8
\end{array} \right)
\begin{pmatrix}
u_0 \\
u_1 \\
u_2 \\
u_{3a} \\
u_{3b} \\
u_{4a} \\
u_{4b} \\
u_5 \\
u_6\\
u_7 \\
u_8
\end{pmatrix} = \begin{pmatrix}
q_0\bar{s}_0 \\
0 \\
-q_0\bar{s}_0 \\
0 \\
p_0\bar{s}_0 \\
0 \\
-p_0\bar{s}_0 \\
0 \\
0 \\
-p_0\bar{s}_0 \\
0 
\end{pmatrix}.\hspace{1mm}
\end{equation}
\ifthenelse{\boolean{ieee}}{\vspace{0.5mm}\\\rule{\textwidth}{0.1 mm}\\\vspace{-12mm}\\}{}
\end{figure*}
Note that the expressions of the $A$ and $B$ matrices used in the model
can be easily derived 
from the first four and last seven rows of the matrix in (\ref{eq-faults2});
just set $p_k=q_k=1$ there (i.e.\ set $P=Q=I$ in (\ref{APBQ})).

\ifthenelse{\boolean{ieee}}{In this model we denote}{Here}
$\bar{s}_0=s_0/(p_0^2+q_0^2)$. 
We emphasize the fact that all the
variables are now homogeneous, and that 
the model takes the symmetric form
(\ref{homogsym}).
Note
only that
excitation terms are ruled out for all branches except for the 
0-th one; that is, 
we are assuming that there are no sources in the remaining branches by
implicitly setting $s_1= \ldots = s_8=0$.
%
%
%
If desired, you can further simplify the model 
by setting 
$p_0=1$, $q_0=0$ (capturing the assumption that the source is an ideal voltage source),
so that $\bar{s}_0=s_0=v_0$
and $u_0$ can be replaced by the current variable $i_0$; 
additionally,
you can set $p_5=1$ and $q_5=z_5$, and replace $u_5$ by the 
current variable $i_5$, since we are not
modelling open-circuits in the fifth branch.
Again, this would lead to a partially homogeneous model.

For 
completeness,
we can check that
the values of $v_5$, computed
as $q_5u_5$ from the solutions of the homogeneous model 
(\ref{eq-faults2})
with $v_0=1$, 
are indeed able to isolate up to
thirteen different faults.
Namely, bridging faults (captured in the vanishing of 
$q_6$, $q_7$, $q_8$) yield, respectively, the values
$-0.178-0.230j$, $0.662$ and $-0.330-0.001j$ for $v_5$.
Analogously, the presence of other short-circuit faults,
defined by the respective vanishing
of $q_1$, $q_2$, $q_{3a}$ (or, equivalently, of $q_{3b}$), 
$q_{4a}$ or $q_{4b}$, 
lead to the values 
$0.332+0.001j$, $-0.664-0.002j$, $-0.331$, 
$0.065+0.199j$ and $0.066-0.198j$. Finally, 
open-circuits in the original impedances, captured in the
vanishing of $p_1$, $p_2$, $p_{3a}$, $p_{3b}$ or $p_{4a}$
(or 
$p_{4b}$), respectively, yield
for $v_5$ the values 
$-0.651-0.002j$, $0.329+0.001j$, 
$0.067-0.198j$, $0.065+0.198j$ and $-0.329-0.002j$.
Recall that in the no-fault case we have $v_5=0$.
In any case, it is worth emphasizing that the interest of the
simulation 
does not rely on the actual values 
met by $v_5$, but on the fact that all computations are performed
in terms of one and the same model (\ref{eq-faults2}), as indicated above.
Even if constructed (for illustrative purposes)
at a small scale,
our example suggests that this approach
may be fruitful in the definition of fault simulation strategies
in larger scale circuits.


\section{Concluding remarks}
\label{sec-con}

The homogeneous formalism introduced in this paper
seems to be of interest in the study of other analytical aspects
of linear circuits.
Further reductions related to nodal analysis,
as well as 
partially homogeneous 
models
and computational aspects,
are the subject of ongoing research. 
Additionally, a
nonlinear version of this approach 
has 
proved feasible and is currently in preparation.
Regarding applications, 
homogeneous and partially homogeneous models are probably 
worth being examined further for 
fault isolation, 
computation of 
Th\'evenin equivalents, analysis of circuits with ideal switches, etc.
The inclusion of fully coupled  and
multiterminal devices 
is in the scope of future research.

\ifthenelse{\boolean{ieee}}{}{\end{document}}

\newpage

\onecolumn

\large

\begin{center}
{\large\bf Author's response}
\end{center}

\vspace{2mm}

\noindent Referees' comments are gratefully acknowledged. I have 
performed a thorough
revision of the manuscript following their suggestions.
The changes are detailed in the response to referees' comments below,
but the main improvements are the following:

\begin{enumerate}
\item Several sections (including the Introduction) have been shortened.
\item Controlled sources can be included in this framework and
there is a new section discussing this (Section V). In particular
this includes a new example which analyzes in the homogeneous
framework a hybrid-$\Pi$ model of small-signal
equivalents of transistors.
\item There is a new subsection (III-F) including several examples which
are simpler than the one at the end (this one 
can be found in current Section VI),
and which are intended to provide the reader with a first clue on how
to use these methods in practice.
\end{enumerate}

\noindent Space restrictions require omitting some details
and prevent a deeper
discussion of certain topics (e.g.\ fully coupled devices beyond controlled
sources) but at least certain hints are given for all these cases.

All reviewers' comments have been addressed, as detailed in what follows. 
Referees' comments are in italics.

\vvv

\noindent {\bf Reviewer 1}

\vv

\noindent {\em 
1.1. The  paper presents an interesting idea of formulating the circuit equations   
in terms of a single abstract variable u, called a homogeneous variable. However, 
this method is limited to the circuits without coupling effects.  In my opinion, this is the basic drawback.}

\vv

\noindent Coupling effects and, in particular, controlled sources can be included
in this framework, as addressed in a new section
(Section V in this manuscript version). Certainly, things get more 
complicated in this context but the main ideas still apply to
circuits with controlled sources. Specifically:

\begin{itemize}
\item homogeneous ($p$, $q$) parameters can still be used, 
providing a comprehensive framework where all four types of controlled
sources (plus switches, etc.) can be accommodated in a unified
manner (pp.\ 11-12);

\vv

\item homogeneous variables $u$ can still be used by means of the 
matrices $\hat{P}$ and $\hat{Q}$ in (\ref{auxcontrolled}); moreover, based on
these matrices, the determinant
characterizing non-degenerate problems
is preserved in the key reduction step as indicated at the end of Section V,
akin to Lemma 1 (p.\ 9) in the uncoupled setting;

\vv

\item tree-based determinantal methods may still be used to characterize degeneracies,
even though there is no space for a detailed analysis of this (see equation (\ref{detCS}) and
the remark regarding control parameters and spanning tree pairs in the last paragraph of Section V);

\vv

\item fully coupled problems can also be included by using appropriate
$P$ and $Q$ matrices; in this setting these matrices would no longer
have a diagonal form (as in the uncoupled context) or a triangular
form (as for controlled sources) and explicit maximal rank conditions, etc.\
seem necessary. It is clear that there is no space
to discuss this in the paper but
I do believe that this line is promising 
e.g.\ in the general analysis of 2-ports and that it 
may provide a nice setting
to address  certain analytical problems in this context in greater generality 
(think e.g.\ of Brune's tests). In my opinion, the
results in previous contexts support the idea that this is an interesting 
research direction.
\end{itemize}

\vv

\noindent {\em  
1.2. I propose, to abridge the sections I  because the text of this section  presents in many places materials well-known. Instead, the author should add a section about the main (at least) ideas how to extend the method to circuits with controlled sources.  The method of formulating  and solving the circuits without controlled sources has a small  practical meaning.}

\noindent Indeed, former section 1 included a (somewhat non-conventional) look at 
some truly elementary facts of circuit theory,
intended to motivate the approach in the paper.
In any case also the third reviewer suggested shortening that section
and I have followed their advice, also in order to get space
for other examples and for the
results involving controlled sources. The (now shorter) motivation
is stated in terms of an example which is revisited in subsection III-F
and at the end of IV-C.
Even if the current example in the first paragraphs of the Introduction is
still a simple one (which is not bad in order to make things
easier to the reader) the tree-based results supporting the 
determinantal expansions arising
there are not trivial. Regarding controlled sources, please check the
answer to item 1.1 above.

\vv

\noindent {\em  
1.3. I propose to extend the section V, also.  The presented example is limited to formulation of the equations, only. And for the completeness it will be very useful to present more information about the method of solving these equations.}  
\vv

\noindent This is the point where I have found more difficulties.
Let me elaborate. On the one hand, the new additions 
(including the examples in III-F, which in a way complement the
one at the end) and the 14-pages
limit make it very difficult to extend this section. 
On the other, and more important, the key aspect of 
the homogeneous formalism is that it provides
an alternative method to set up the circuit models, and for this
reason the emphasis is on the form that the different equations take in
practice. In the example of
Section VI (former Section V) we arrive at linear
systems of order eleven which can solved automatically in a
straightforward manner (I simply used the predefined solver in Maple).

It is true, however, that in larger scale systems this should be 
accompanied by a discussion on how to set up the $A$ and $B$ matrices
(typically they can be computed from a spanning tree: cf.\ (\ref{fundamental})
and also the remark at the end of the fourth paragraph
in the second column of p.\ 13 about the form of these
matrices in the example; I must say that
simply including another 11x11 matrix
in the manuscript would consume a substantial amount of space)
and also by numerical linear algebra aspects (LU decompositions, etc.).
Unfortunately I don't find space for this but on the other hand
I believe that it can be skipped for a relatively small scale example. 

In this regard, and also for eventual large scale applications 
of this approach, this formalism should be extended to
other techniques used in CAD tools, 
particularly to nodal analysis. This extension is feasible,
essentially along the lines in which Modified Nodal Analysis (MNA) extends
basic nodal analysis; the key idea is to use additional (homogeneous)
variables
only for the branches that for whatever reason need 
a homogeneous description, and
a new example (namely, (\ref{MNA})) has been added to illustrate this.
In greater detail, these models require a deeper discussion and this topic
is in the scope of future research.

\vvv

\noindent {\bf Reviewer 2}

\vv

\noindent {\em  2.1. In this paper, the approach to the description and the analysis of linear circuits based on elementary aspects of projective geometry is developed. The proposed method is interesting, but the practical contribution of the research is not clear. Only one example of fault diagnosis of the passive circuit (Wien bridge) is presented.}

\vv

\noindent Essentially, the paper develops a theoretical idea (modelling
devices and circuits in a homogeneous manner); I believe that it has
many potential applications but for space restrictions there is no chance
to discuss them in detail. 
To support this claim, let me mention fault diagnosis problems
(the Wien bridge in Section VI is only an example but the idea can be used 
further) but also 
for instance 
the computation of Th\'evenin equivalents (an
example 
is given in subsection III-F)
or problems
involving controlled sources, switches, etc. which are now 
included
in Section V. Broadly speaking, the approach should be useful
in theoretical or real problems in which
one needs to model the full spectrum of parameter values for certain 
(or 
all) devices. In any case I assume that the work
has essentially a theoretical value
and that future work developing
further computational aspects, the hybridization with nodal-based CAD 
techniques, etc. will possibly extend the scope of 
applications of this approach.

\vv

\noindent {\em  2.2. The analysis of circuits with coupling effects, multiterminal devices, etc. is not discussed.}

\vv

\noindent Controlled sources are now included. Fully coupled problems and 
multiterminal devices are left for future work; they can
be accommodated in this framework but possibly with additional 
difficulties. Please find further details in the answer to items 1.1 and 2.7.

\vv

\noindent {\em  2.3. The analysis of modern circuit is unreasonable 
without CAD tools. But in this paper, the algorithm available for 
implementation in a computer program is not presented. The computational 
efficiency of the proposed technique is also not clear. The term 
cancellation problem which arises from the usage of the matrix 
formulation in analog circuits is not mentioned.}

\vv

\noindent Please see my answer to item 1.3 above. In the 
current version
 I mention
some of the difficulties (including the term cancellation problem) 
that may arise in more general contexts,
in particular in fully coupled problems,
at the end of Section V (page 12).

\vv

\noindent {\em 2.4. In view of this, the authors may take note of the following: 

\vv

\noindent (Item 1 in the reviewers' report) This reviewer would suggest enhancing the discussion of practical use and benefits of the proposed method. The authors should compare with other methods used in modern circuit design.}

\vv

\noindent The homogeneous approach is somehow transversal to the different
analysis methods in circuit design, in the sense that 
homogeneous descriptions can be used in the context of branch-oriented
methods but also in nodal analysis, etc. The examples in
the new subsection III-F 
are intended to illustrate this. The main advantage of this
framework is a theoretical one but also, in practice, by means of this approach
we
may accommodate in a single model and simultaneously all possible
parameter values for the circuit devices,
including short- and open-circuits (note that 
in the classical framework,
handling both configuration usually requires setting
up two different models).
In any case future work should address a systematic analysis of how
to extend Modified Nodal Analysis (MNA) techniques, used in many
circuit simulators, to accommodate homogeneous descriptions
in some branches; the example (\ref{MNA})
briefly suggests how to do this.
Please check also my answer to item 2.1 regarding the practical use of this 
approach.

\vv

\noindent {\em 2.5 (item 2 in the report). The paper is rather long. The authors should consider reducing some parts.}

\vv

\noindent Following also the suggestion of the other reviewers I
have reduced substantially the Introduction and, to a lesser extent,
other sections (in particular subsection II-E). But on the other hand
the inclusion of controlled sources and additional examples (as suggested
in the reports) still makes the paper 14-pages long.

\vv

\noindent {\em 2.6 (item 3 in the report). The explanation of presented example is not clear enough. It is better to use some step by step algorithm.}

\vv

\noindent Please see my answer to item 1.3 above. With the space
restrictions and the suggestions for additions I have found it better
to detail the way in which the models are set up by means of 
additional (simpler) examples which can be now found on Sections III-F and V.

\vv

\noindent {\em 2.7 (item 4 in the report). The benefits of modern circuit analysis method are rather low if the depended sources, nullors, and pathological mirrors are not allowed. The authors should consider extending the method to multiterminal models of active devices.} 

\vv

\noindent Controlled sources are now included (Section V);
please find a discussion of the benefits of our formalism in this
context in the response to item 1.1 above.
Multiterminal devices and fully coupled problems
are 
research lines in the scope of future work, even if
the current manuscript version now includes a transistor example
(Section V). Worth mentioning
is the fact that the chance to use homogeneous descriptions only
for certain devices (as in the partially homogeneous models within 
sections III-F and VI) also provides a way to combine this formalism
with classical methods accommodating multiterminal devices and so on.

\vvvv

\noindent {\bf Reviewer 3}

\vv

\noindent {\em  3.1. The paper uses ideas and techniques from projective geometry for
analyzing linear circuits. It is shown that these techniques make
it possible to avoid in a first phase the typical assumptions of voltage or current controlled two-terminal elements used for deriving the formulation of reduced models. The paper is carefully written and technically sound. It appears
the devised techniques provide benefits in certain aspects of 
linear circuit analysis. I have the following main comments. 

1. Some parts of the paper are maybe too descriptive and also
repetitive.}

\vv

\noindent I have substantially shortened Section I (Introduction), 
where some of the material was in a way repeated later. Additionally, 
it is true that subsections II-B and II-C somehow go in parallel, but
this is not by chance (and this is explicitly acknowledged at the beginning
of II-D). Both contexts are of independent interest (impedances in II-B, 
sources in II-C) and from my point of view it is important to give details
for both of them. Additionally, I performed a thorough reading
of the manuscript and deleted some remarks which were indeed repeated. 
Thank you.

\vv

\noindent {\em 3.2.
Some concepts of projective geometry, although they
are probably elementary for a specialist, are not so easy to
follow for a non-specialist.}

\vv

\noindent Sure, this is not an easy topic. For this reason
a detailed introduction is given in II-A. Hopefully the discussion
regarding the voltage-controlled (admittance) and current-controlled
(impedance) patches later in subsections II-B and III-B, much
closer to the circuit-theoretic language, will
be of help in this regard for many TCAS readers.

\vv

\noindent {\em  3.3. In the current version there are
elementary examples in the introduction to explain the basic
ideas and a much more elaborated example in Section V. I feel
it would be useful to discuss at least one addition example
of intermediate complexity to illustrate the benefits of the
method. If needed some parts of the paper can be shortened
avoiding repetitions or too extended discussions.}

\vv

\noindent Please find such examples in the new 
subsection III-F. Indeed
I had to shorten other parts to include this and the 
material on controlled sources (Section V) in order to comply with the
14-pages limit.

\vv

\noindent {\em 3.4 (item 2 in the reviewer's report).
The proposed approach has advantages but, as stated by the
Authors, it lacks an immediate physical meaning, due to the use 
of additional variables. Discussing additional examples (cf.
Point 1} [item 3.3]{\em ) may help the reader to re-gain some physical insight.}

\vv

\noindent Yes. New examples are now included in III-F. Additionally,
the changes in the Introduction should help the reader understand
better the meaning of the $u$ variables (and of the homogeneous
parameters $p$, $q$) in the context of the parametric
form of Ohm's law, namely $i=pu,$ $v=qu$.

\vv

\noindent {\em 3.5 (item 3 in the report). 
The main result in Theorem 1 is closely related to previous
results by Chen obtained via a different method. A brief
specific comparison with those results is given after (44).
The Author might want to add some further comments comparing
from a more general viewpoint Chen approach and that in the
manuscript.}

\vv

\noindent Many thanks. For me and from a theoretical point
of view this is one of the most important points of the paper.
Besides the remarks given after (44) ((\ref{mmt0}) in the
current version, page 10), I believe that the main difference
comes from the fact that, in weighted settings, Chen always needs
to resort to an impedance or admittance or (eventually) hybrid
description for the set of devices, that is, 
to fix either a current-controlled or a voltage-controlled description 
for every single branch; by contrast, 
all our results in Section IV benefit
from the homogeneous formalism by keeping the matrices $P$ and $Q$
throughout. In other terms, we can state the
(weighted) matrix-tree theorem not necessarily
in terms of the nodal-admittance or the loop-impedance matrices,
but in terms of the (say) homogeneous version of both matrices 
which is depicted
in (\ref{APBQ}). I try to explain this (briefly, again because
of space restrictions) at the beginning
of Section 4. Additionally, 
more on this can be found right after the proof of Proposition
1 (second paragraph on page 10).

\vv

\noindent {\em 3.6 (item 4 in the report). The introduction appears to be quite long. Is there a way to
better organize it in order to more neatly pinpoint the 
motivations, background and main results?}

\vv

\noindent I have changed and substantially shortened Section I. 
Please check also the answer to item 1.2 above.

\


\vvv

Many thanks. Best regards,


Ricardo Riaza

\end{document}